\documentclass[preprint,aps,superscriptaddress,showpacs,nofootinbib]{revtex4-1}
\usepackage{geometry}

\usepackage{amscd}
\usepackage{amsmath,amssymb,amsthm,mathrsfs,amsfonts,dsfont}
\usepackage{subfigure, epsfig}
\usepackage{braket}
\usepackage{bm}
\usepackage{enumerate}

\usepackage{graphicx}
\usepackage{epsfig}
\usepackage{subfigure}
\usepackage{color}
\usepackage[10pt]{moresize}

\newcommand\oprod[2]{\ensuremath{|#1\rangle\langle#2|}}
\newcommand\mean[1]{\ensuremath{\langle #1 \rangle}}

\begin{document}

\title{Experimental Twin-Field Quantum Key Distribution Over 1000 km Fiber Distance}

\author{Yang Liu}
\affiliation{Hefei National Research Center for Physical Sciences at the Microscale and School of Physical Sciences, University of Science and Technology of China, Hefei 230026, China}
\affiliation{Jinan Institute of Quantum Technology and CAS Center for Excellence in Quantum Information and Quantum Physics, University of Science and Technology of China, Jinan 250101, China}
\affiliation{Hefei National Laboratory, University of Science and Technology of China, Hefei 230088, China}

\author{Wei-Jun Zhang}
\affiliation{State Key Laboratory of Functional Materials for Informatics, Shanghai Institute of Microsystem and Information Technology, Chinese Academy of Sciences, Shanghai 200050, China}

\author{Cong Jiang}
\affiliation{Jinan Institute of Quantum Technology and CAS Center for Excellence in Quantum Information and Quantum Physics, University of Science and Technology of China, Jinan 250101, China}
\affiliation{Hefei National Laboratory, University of Science and Technology of China, Hefei 230088, China}

\author{Jiu-Peng Chen}
\author{Chi Zhang}
\affiliation{Hefei National Research Center for Physical Sciences at the Microscale and School of Physical Sciences, University of Science and Technology of China, Hefei 230026, China}
\affiliation{Jinan Institute of Quantum Technology and CAS Center for Excellence in Quantum Information and Quantum Physics, University of Science and Technology of China, Jinan 250101, China}

\author{Wen-Xin Pan}
\affiliation{Hefei National Research Center for Physical Sciences at the Microscale and School of Physical Sciences, University of Science and Technology of China, Hefei 230026, China}

\author{Di Ma}
\affiliation{Jinan Institute of Quantum Technology and CAS Center for Excellence in Quantum Information and Quantum Physics, University of Science and Technology of China, Jinan 250101, China}

\author{Hao Dong}
\affiliation{Hefei National Research Center for Physical Sciences at the Microscale and School of Physical Sciences, University of Science and Technology of China, Hefei 230026, China}
\affiliation{Jinan Institute of Quantum Technology and CAS Center for Excellence in Quantum Information and Quantum Physics, University of Science and Technology of China, Jinan 250101, China}

\author{Jia-Min Xiong}
\affiliation{State Key Laboratory of Functional Materials for Informatics, Shanghai Institute of Microsystem and Information Technology, Chinese Academy of Sciences, Shanghai 200050, China}

\author{Cheng-Jun Zhang}
\affiliation{Photon Technology (Zhejiang) Co. Ltd., Jiaxing 314100, China}

\author{Hao Li}
\affiliation{State Key Laboratory of Functional Materials for Informatics, Shanghai Institute of Microsystem and Information Technology, Chinese Academy of Sciences, Shanghai 200050, China}

\author{Rui-Chun Wang}
\author{Jun Wu}
\affiliation{State Key Laboratory of Optical Fibre and Cable Manufacture Technology, Yangtze Optical Fibre and Cable Joint Stock Limited Company, Wuhan, 430073, China}

\author{Teng-Yun Chen}
\affiliation{Hefei National Research Center for Physical Sciences at the Microscale and School of Physical Sciences, University of Science and Technology of China, Hefei 230026, China}
\affiliation{Hefei National Laboratory, University of Science and Technology of China, Hefei 230088, China}

\author{Lixing You}
\affiliation{State Key Laboratory of Functional Materials for Informatics, Shanghai Institute of Microsystem and Information Technology, Chinese Academy of Sciences, Shanghai 200050, China}

\author{Xiang-Bin Wang}
\affiliation{Jinan Institute of Quantum Technology and CAS Center for Excellence in Quantum Information and Quantum Physics, University of Science and Technology of China, Jinan 250101, China}
\affiliation{Hefei National Laboratory, University of Science and Technology of China, Hefei 230088, China}
\affiliation{State Key Laboratory of Low Dimensional Quantum Physics, Department of Physics, Tsinghua University, Beijing 100084, China}

\author{Qiang Zhang}
\affiliation{Hefei National Research Center for Physical Sciences at the Microscale and School of Physical Sciences, University of Science and Technology of China, Hefei 230026, China}
\affiliation{Jinan Institute of Quantum Technology and CAS Center for Excellence in Quantum Information and Quantum Physics, University of Science and Technology of China, Jinan 250101, China}
\affiliation{Hefei National Laboratory, University of Science and Technology of China, Hefei 230088, China}

\author{Jian-Wei Pan}
\affiliation{Hefei National Research Center for Physical Sciences at the Microscale and School of Physical Sciences, University of Science and Technology of China, Hefei 230026, China}
\affiliation{Hefei National Laboratory, University of Science and Technology of China, Hefei 230088, China}

\begin{abstract}
Quantum key distribution (QKD) aims to generate secure private keys shared by two remote parties. With its security being protected by principles of quantum mechanics, some technology challenges remain towards practical application of QKD. The major one is the distance limit, which is caused by the fact that a quantum signal cannot be amplified while the channel loss is exponential with the distance for photon transmission in optical fiber. Here using the 3-intensity sending-or-not-sending protocol with the actively-odd-parity-pairing method, we demonstrate a fiber-based twin-field QKD over 1002 km. In our experiment, we developed a dual-band phase estimation and ultra-low noise superconducting nanowire single-photon detectors to suppress the system noise to around 0.02 Hz. The secure key rate is $9.53\times10^{-12}$ per pulse through 1002 km fiber in the asymptotic regime, and $8.75\times10^{-12}$ per pulse at 952 km considering the finite size effect. Our work constitutes a critical step towards the future large-scale quantum network.
\end{abstract}

\maketitle

{\it Introduction.---}
With the security protected by laws of quantum mechanics, quantum key distribution (QKD)~\cite{bennett1984quantum,ekert1991quantum,gisin2002quantum,scarani2009security,gisin2015far,xu2020secure,pirandola2020advances} can distribute secret private keys between remote parties through photon transmission. Given the advantage in security, there are also some barriers to the practical application of QKD. The biggest challenge is the channel loss to the single-photon level weak light used in QKD. This limits the secure distance  of practical QKD severely since a quantum signal cannot be perfectly cloned~\cite{wootters1982noncloning}. The channel transmittance decreases exponentially with the distance, hence the channel and detection noise prevents the QKD system to produce secure key in a long distance. Moreover, with the number of heralded events being smaller in longer distance QKD, the finite-key effect becomes a problem that reduces the generation of key bits. Through the extensive studies of QKD in the past decades, much progress has been made towards practical applications. Alongside these studies, the secure distance has been raised drastically with several notable breakthroughs in both theory and technology, especially after the twin-field QKD (TF-QKD) is proposed~\cite{lucamarini2018overcoming}.

The key rate of traditional QKD protocols scales linearly with the channel transmittance $\eta$. TF-QKD~\cite{lucamarini2018overcoming} improves this relation to $\sqrt\eta$ without using a quantum memory or trusted relay. This provides a promising way to a longer secure distance of point-to-point QKD and to a large-scale quantum network with fewer trusted relays. So far, TF-QKD has been demonstrated experimentally in lab~\cite{minder2019experimental,wang2019beating,liu2019exp,zhong2019proof,fang2020implementation,chen2020sending,liu2021field,chen2021twin,chen2022quantum,pittaluga2021600,wang2022twin} through up to 830-km spooled fiber~\cite{wang2022twin}, and in the field test through 511-km deployed fiber between metropolitans~\cite{chen2021twin}. 

Here we demonstrate a TF-QKD using sending-or-not-sending (SNS) protocol~\cite{wang2018sns} over long distance fiber spools. The advanced 3-intensity decoy-state method~\cite{hu2022universal} and the actively-odd-parity-pairing (AOPP)~\cite{xu2020sending, jiang2020zigzag} are used to improve the key rate. The system noise is suppressed in experiment using narrow filter assisted superconducting nanowire single-photon detectors (SNSPDs) and using a data post-processing based dual-band phase estimation method. We achieve a 1002 km distribution distance in the asymptotic regime, and a 952 km distribution distance considering the finite size effect. Further, a 47.06 kbps secure key rate is achieved through 202 km fibers.

\begin{figure*}[tbh]
\centering
\resizebox{14cm}{!}
{\includegraphics{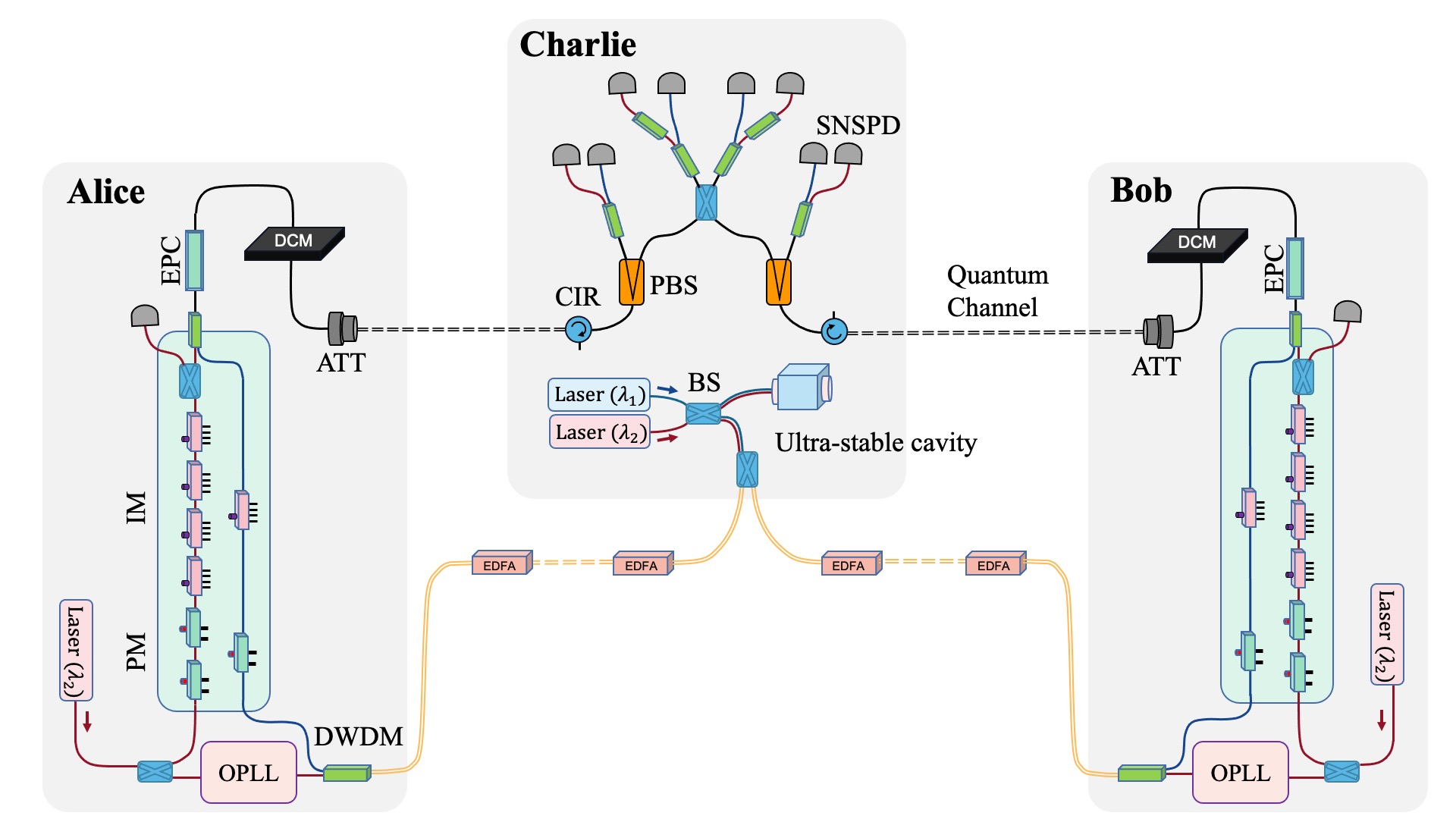}}
\caption{Experimental Setup.
In Charlie's station, two lasers ($\lambda_1$=1548.51 nm and $\lambda_2$=1550.12 nm) are frequency locked to an ultra-stable cavity, then combined, and distributed to Alice (Bob) via 450 km (450 km) single mode fiber spools.
In Alice's (Bob's) station, $\lambda_2$ is regenerated with an optical phase-locked loop (OPLL). The $\lambda_2$ light is encoded to ``dim phase reference'' and quantum signal with time-multiplexing, then combined with the $\lambda_1$ ``strong phase reference'' with wavelength-multiplexing. Signals from Alice and Bob are transmitted to Charlie for interference. The interference results are measured with SNSPDs. Additional SNSPDs are used to monitor the signal intensities at Alice and Bob, the polarization and the relative delay at Charlie.
BS, beam splitter; PBS, polarization beam splitters; IM, intensity modulator, PM, phase modulator; ATT, attenuator; DWDM, dense wavelength division multiplexing; CIR, optical circulator; EDFA, erbium-doped fibre amplifier; DCM, dispersion compensation module; EPC, electronic polarization controller.
}
\label{Fig:TFQKD_Setup}
\end{figure*}

{\it Protocol.---}
We adopt the SNS protocol~\cite{wang2018sns} using 3 intensities (3-intensity SNS-TF-QKD) and the improved key rate calculation~\cite{hu2022universal}. We also apply the AOPP~\cite{xu2020sending} method to reduce the bit-flip error rate. Different from ~\cite{yu2019sending, jiang2019unconditional}, we use the advanced decoy-state analysis~\cite{hu2022universal} to improve the secure key rate. Say, Alice takes decoy-state analysis after bit-flip error correction. This means we can use the whole numbers of heralded time windows of each intensity for the decoy-state analysis~\cite{hu2022universal}. This improves the key rate because on the one hand, we don't need to spend a subset of code bits only for decoy-state analysis, on the other hand, we have a larger data size in the decoy-state analysis and hence the finite-data effect is reduced. There are three weak coherent state sources from Alice and Bob: the vacuum source $v$, the decoy source $x$, and the signal source $y$ whose intensities are $\mu_v=0,\mu_x,\mu_y$ respectively. In each time window, Alice (Bob) randomly chooses one source $l$ ($r$) from the three candidate sources with probability $p_l$ ($p_r$) with $l,r=v,x,y$ and then sends the prepared pulse to the measurement station which is controlled by Charlie. Charlie is assumed to measure the interference result of the incoming pulse pair and announce the measurement results to Alice and Bob. In the protocol, Alice and Bob send $N$ pulse pairs to Charlie, then Charlie announces those time windows with correct heralding. Finally, they distill the secure keys according to the following formula:
\begin{equation}
R=\frac{1}{N}\{n_1^\prime[1-H(e_1^{\prime \mbox{ph}})]-fn_t^\prime H(E_t^\prime)\},
\label{Eq:KeyRate}
\end{equation}
where $R$ is the key rate of per sending-out pulse pair; $n_1^\prime$ is the lower bound of the number of survived untagged bits after AOPP and  $e_1^{\prime \mbox{ph}}$ is the upper bound of the phase-flip error rate of those survived untagged bits after AOPP; $n_t^\prime$ is the number of survived bits after AOPP and $E_t^\prime$ is the corresponding bit-flip error rate in those survived bits; $f$ is the error correction inefficiency which we set to $f=1.16$; $H(x)=-x\log_2x-(1-x)\log_2(1-x)$ is the Shannon entropy. To consider finite-key effects, they need to estimate parameters $n_1^\prime$ and $e_1^{\prime\mbox{ph}}$ faithfully with finite data, also, tailing terms~\cite{hu2022universal} have to be added to Eq.~(\ref{Eq:KeyRate}) \cite{jiang2020zigzag,jiang2021composable,hu2022universal}. (See Supplemental Materials for details of the theoretical calculations).

{\it Experiment.---}

The experimental setup is shown in Fig.~\ref{Fig:TFQKD_Setup}. Two lasers with wavelengths $\lambda_1$ (1548.51 nm) and $\lambda_2$ (1550.12 nm) are frequency locked to an ultra-stable cavity at Charlie, using the Pound-Drever-Hall (PDH) technique~\cite{pound1946electronic,drever1983laser}. The Hertz level line-width light is then sent to Alice and Bob.

Alice and Bob modulate the $\lambda_1$ light from Charlie to a 400 ns pulse in each 1 $\mu$s period, as the ``strong phase reference''. A locally prepared laser with a nominal line-width of 1 kHz is locked to the $\lambda_2$ light from Charlie using an optical phase-locked loop (OPLL). This $\lambda_2$ light is modulated with a 100 ms period. The initial 40 ms of light serves as the ``dim phase reference'', with intensity set to a higher level for phase estimation at Charlie; the remaining 60 ms is used as ``quantum signals'', with intensity set to the single photon level. Four different relative phases between Alice and Bob ($\delta_{AB}=\{0, \pi/2, \pi, 3\pi/2\}$) are modulated for both the ``strong phase reference'' and ``dim phase reference''.

The quantum signals are modulated to 3 intensities and 16 phases, following the 3-intensity SNS-TF-QKD protocol. A fraction of the $\lambda_2$ light is monitored to stabilize the intensities of the decoy states and the ``dim phase reference'', before attenuating to a single photon level. Then, the $\lambda_1$ and $\lambda_2$ light are combined and transmitted to Charlie through ultra-low-loss fibers with an average attenuation of less than 0.157 dB/km. 

At Charlie's measurement station, one output of the PBS is used to monitor the light from Alice and Bob. The polarization is adjusted according to the detection rate at this monitoring port. The $\lambda_1$ detections is set between 75 kHz and 300 kHz, and the $\lambda_2$ as low as possible. The rising edges of the $\lambda_1$ pulses are used to compensate the relative delay between Alice's and Bob's signals (See Supplemental Materials for details of the feedback system). The light from the signal ports of the PBSs are combined for interference. The interference output of the $\lambda_1$ and $\lambda_2$ light are separated and filtered by DWDMs, measured with SNSPDs, and recorded with a Time Tagger.

The most challenging barriers to achieving ultra-long distance TF-QKD lies in reducing the channel loss and the system noise. To reduce channel loss, we adopted the ``pure silica core'' technology by reducing the doped Ge in the core, and we decreased the fictive temperature in the manufacturing process. We adopted a large effective area design with around 125 $\mu m^2$ effective area, to reduce the non-linear effect in transmission. To reduce the system noise, we adopted low temperature filters in the SNSPDs resulting a lower dark count rate (DCR); we developed a time-multiplexing dual-band phase estimation method that helps to control spurious photons from the channel.

The main contribution of the dark count noise of the NbN SNSPDs is the blackbody radiation coming along the input fiber. This noise is reduced using multiple filters. The long-wavelength ($>$2 $\mu$m) noise photons are filtered by coiling the fiber to a 28 mm diameter at 40 K cold plate. Other blackbody photons are filtered out by inserting a customized cryogenic bandpass filter (BPF) before coupling the fiber to the SNSPD chip at 2.2 K cold plate~\cite{zhang2018fiber}. The BPF is centered at 1550 nm with a 5 nm bandwidth and a 85\% transmittance. After suppressing the blackbody photons, the DCR is reduced from about 10 Hz to about 0.02 Hz. Further, the photon absorption is optimized by using a distributed Bragg reflector (DBR) based optical cavity~\cite{zhang2017nbn} to enhance the detection efficiency to 94 $\pm$ 2\%. We note the system detection efficiency decreases to around 60\% due to the loss in transmitting and insertion loss of the BPF.

The dominant source of noises from the channel is identified as the re-Rayleigh scattering of the phase estimation signal in pervious studies~\cite{chen2020sending,chen2021twin}. For example, approximately 8 Hz re-Rayleigh scattering is expected when the detected rate of the time-multiplexed phase reference is 2 MHz~\cite{chen2020sending,chen2021twin}. The dual-band stabilization method reduces the re-Rayleigh scattering noise by setting the wavelength of the phase reference light different from the quantum signal~\cite{pittaluga2021600}. The re-Rayleigh scattering generated by the $\lambda_1$ ``strong phase reference'' can be easily filtered with a DWDM. After compensating the fast phase fluctuation with $\lambda_1$, the residual phase of the $\lambda_2$ quantum signal can be corrected with a much weaker ``dim phase reference'' time-multiplexed with the quantum signal.

However, in the ultra-long fiber scenario, the noise induced by the $\lambda_1$ ``strong phase reference'' cannot be neglected. The measured noises in the $\lambda_2$ channel reach approximately 0.55 Hz when the detected reference count is 2 MHz through a 300 km fiber. As a comparison, the crosstalk noise to the $\lambda_2$ channel is around $10^{-3}$ Hz when removing the fiber spools and setting the same $\lambda_1$ detection rate. We attribute the main source of noise to the spontaneous Raman scattering of the $\lambda_1$ light as it propagates through the fiber. (See Supplemental Materials for details of the measurement of noises).

To avoid the spontaneous Raman scattering noise, we modulate the ``strong phase reference'' to a 400 ns pulse in the 1 $\mu$s period. The quantum signals is transmitted solely during the last 600 ns when $\lambda_1$ is set to vacuum. The noise in the signal period is filtered with time division.

\begin{figure}[tbh]
\centering
\resizebox{8.5cm}{!}
{\includegraphics{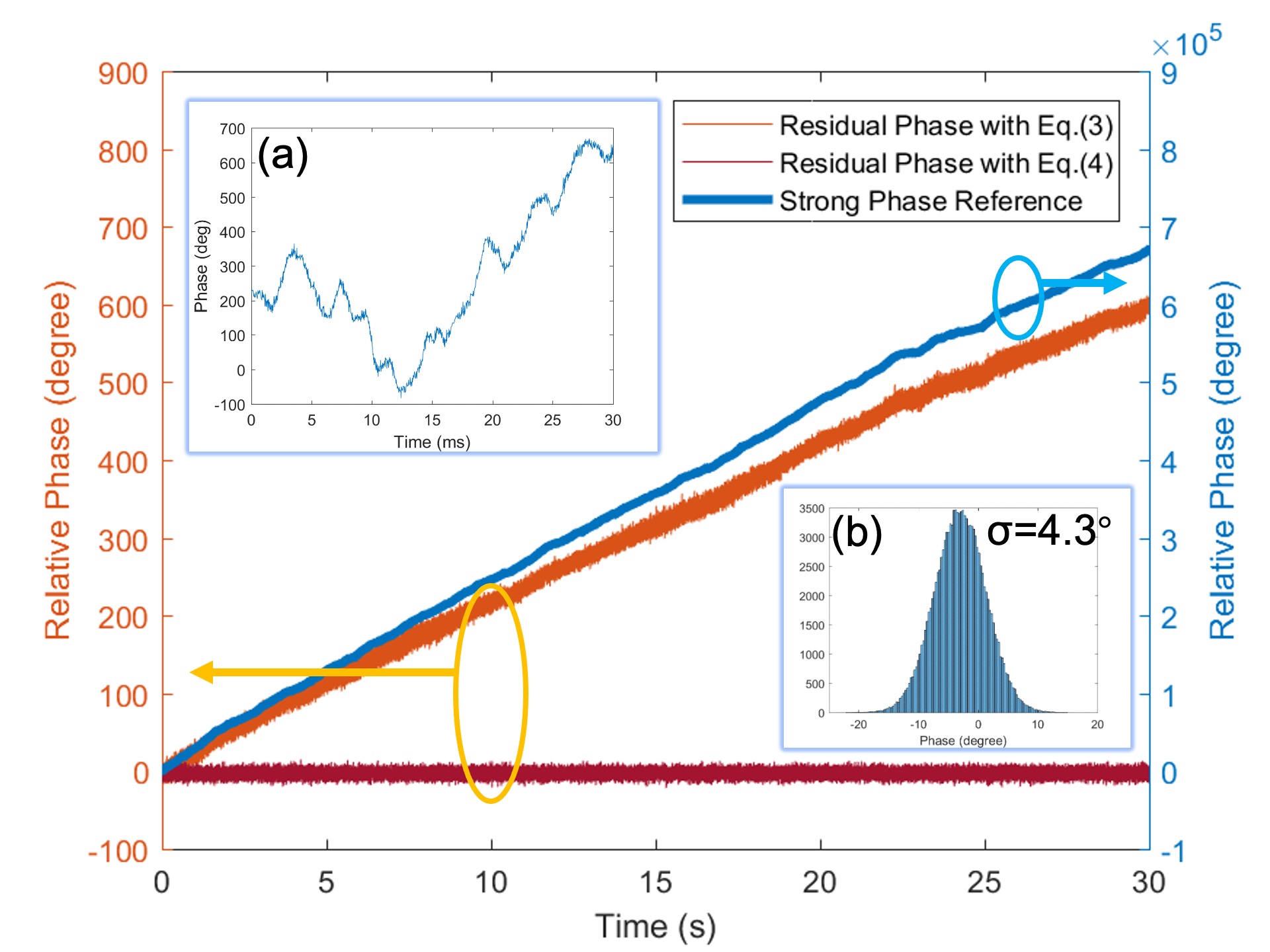}}
\caption{Dual-band stabilization with data post-processing. In the main figure, Blue: Free phase drift estimated with ``strong phase reference'' in 30 s period; Orange: Residual phase drift after simple compensation with Eq. (\ref{Eq:SCorrection}); Red: Residual phase drift after fine compensation with Eq. (\ref{Eq:FCorrection}); Insert (a) The first 30 ms of the free phase drift; (b) Histogram of the residual phase with fine compensation (Red line in main figure). }
\label{Fig:PhaseDrift}
\end{figure}

We developed data post-processing method for the time-multiplexed dual-band phase estimation that removes the need of active phase modulators and real-time feedback circuits for phase compensation.

The first step is to estimate the relative phase of $\lambda_1$ ``strong phase reference''. We follow the procedure in our previous phase estimation method~\cite{liu2019exp} to correct fast phase changes. As shown in Fig.~\ref{Fig:PhaseDrift} and its insert~(a), the phase drift of the ``strong phase reference'' shows a monodirectional drift in a large time scale, while fast oscillations appears in short period.

The second step is to estimate the phase of the quantum signal. The simplest method is to set the ``strong phase reference'' directly as the estimation of the phase of quantum signal:
\begin{equation}
	\phi_s(t) = \phi_r(t)  + \phi_s(0)-\phi_r(0)
	\label{Eq:SCorrection}
\end{equation}

We define the residual phase $\phi_{residual}$ as the difference between the estimated phase $\phi_s(t)$ and the actual phase $\phi_{s\_act}(t)$ of the quantum signal:

\begin{equation}
	\phi_{residual}(t) = \phi_{s\_act}(t)-\phi_s(t)
\end{equation}

The residual phase $\phi_{residual}$ stands for the error induced in the phase estimation process, which will affect the QBER in X basis. A smaller residual phase indicates a better estimation.

With the simple estimation method, the residual phase is already reduced by more than 1,000 times compared with free dirft, similar to the reported hardware-based dual-band compensation~\cite{chen2022quantum}. The residual phase in Eq.(\ref{Eq:SCorrection}) is mainly contributed by the wavelength difference. It is possible to take this effect into consideration in data processing:
\begin{equation}
	\phi_s(t) = \phi_r(t) \times \lambda_2/\lambda_1 + \phi_s(0)-\phi_r(0)
	\label{Eq:FCorrection}
\end{equation}

Taking the wavelength difference into consideration by Eq.~(\ref{Eq:FCorrection}), the residual phase is further suppressed, resulting in a standard deviation of $4.3^\circ$ in the 30 s test. 

The final step is to determine the phase difference $\phi_s(0)-\phi_r(0)$ present in Eq.(\ref{Eq:SCorrection}) and Eq.(\ref{Eq:FCorrection}). Here we optimize this value based on the least squares method, taking both the detections of ``strong phase reference'' and ``dim phase reference'' into account. (See Supplemental Materials for details of the phase estimation). In the experiment, this phase difference is calculated and refreshed every 500 ms, to avoid any accumulated errors due to inaccurate wavelength settings, high order residual phase errors, and accumulated errors in estimating ``strong phase reference''.

{\it Result.---}
Our system runs at a 1 GHz frequency with a signal pulse width of 120 ps. The quantum signals are not sending in the first 400 ns of the 1 $\mu$s period when ``strong phase reference'' emits, or in the first 40 ms of the 100 ms period when ``dim phase reference'' is sending. Further, the detected quantum signals near the edges of these strong light are also dropped, to avoid potential noises. The effective signal frequency is 351 MHz. 

The dark counts of the ultra-low noise SNSPDs are measured to be 0.014 Hz and 0.026 Hz on installation, with detection efficiencies of 60\% and 55\%. The total noises of the $\lambda_2$ quantum channel is measured to about 0.019 Hz and 0.035 Hz, when the ``strong phase reference'' is on. In data processing, we employ a 200 ps window to further filter the noises, whose efficiency is about 65\%. We note the SNSPD dark counts increased during the experimental tests.

\begin{figure}[tbh]
\centering
\resizebox{8.5 cm}{!}
{\includegraphics{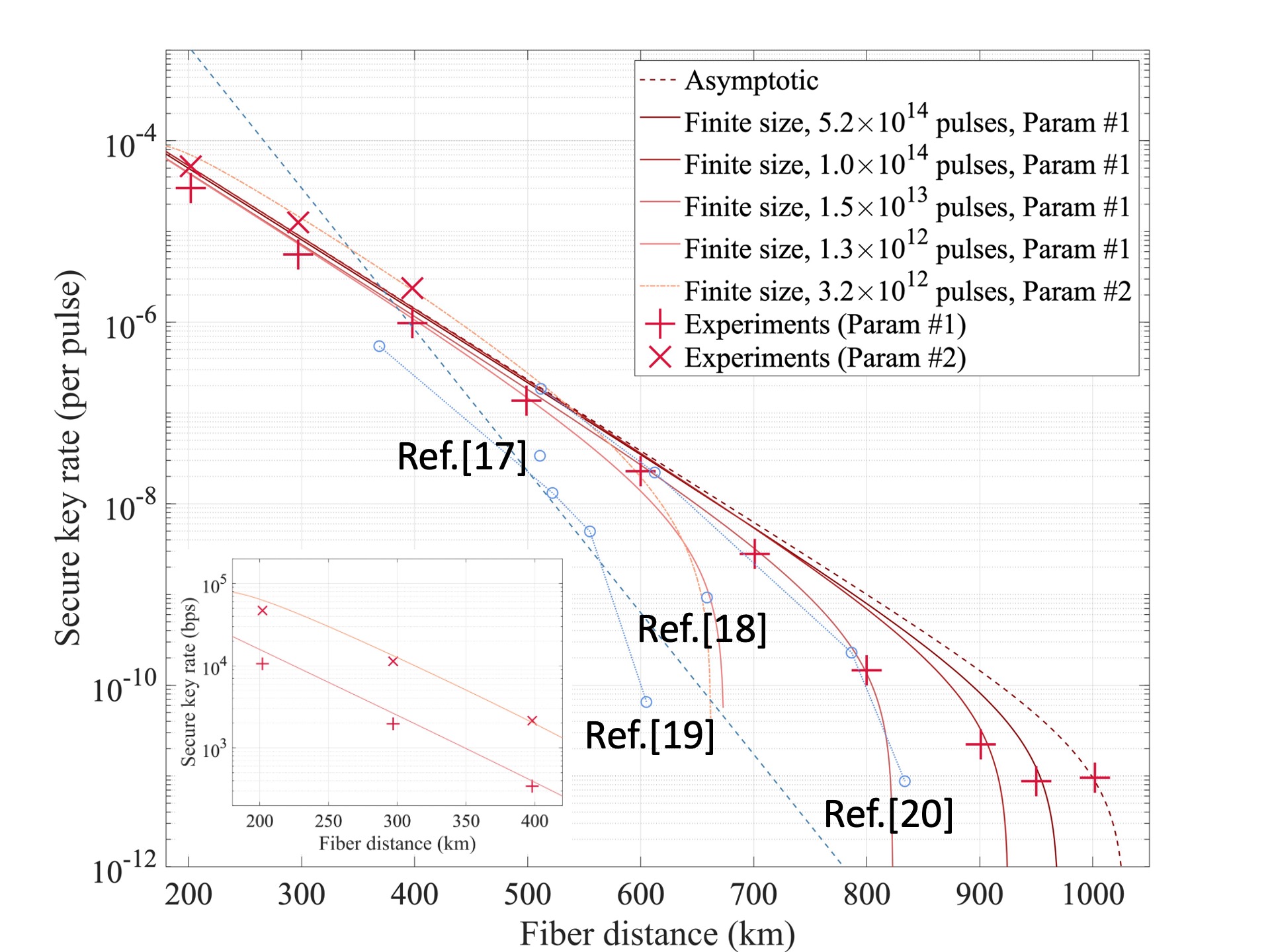}}
\caption{Simulations and experimental results of the secure key rates. The ``$+$''-shape points are our experimental results using the parameters optimized for long distance (Parameter \#1), the ``$\times$''-shape points are our experimental results using the parameters optimized for short distance (Parameter \#2). The solid curves are the simulation results considering the finite size effect. The red dashed curve is the simulation result with infinite size assumption. The circle markers indicates the state-of-the-art TF-QKD results reported in Ref.~\cite{chen2021twin, pittaluga2021600, chen2022quantum, wang2022twin}. The blue dashed line shows the PLOB bound. Insert: the secure key rate per second in short distances.}
\label{Fig:KeyRate}
\end{figure}

We take measurements at total fiber distances between 202 km and 1002 km between Alice and Bob, from 31.6 dB to 156.5 dB in a symmetric fiber setup. The secure key rates are summarized in Fig.~\ref{Fig:KeyRate}. Positive secure keys are generated in asymptotic regime for the 1002 km case, and in finite size regimes for all other distances.

In the longest distribution distance of 1002 km, a total of $1.02\times10^{14}$ signal pulses are sent. The secure key rate is calculated as $9.53\times10^{-12}$, which is 0.0034 bps, considering the effective signal frequency. We note that due to the narrow line-width laser used in the experiment, the fiber length between Alice-Charlie and Bob-Charlie do not have to be exactly the same. For example, the fiber distances in this test are 500 km and 502 km between Alice-Charlie and Bob-Charlie, respectively.

The finite size effect~\cite{jiang2021composable} is taken into consideration for all other experimental tests with the fiber lengths between 202 km and 952 km for composable security under any coherent attack~\cite{jiang2019unconditional, jiang2021composable}. In the calculation, the error correction inefficiency is set to $f=1.16$, the statistics block size is set to the complete length of the test, the failure probability of applying the Chernoff bound in the finite-size estimation is set to $\varepsilon=10^{-10}$, the failure probability of the error correction process and the privacy amplification process are set to $\varepsilon_{cor}=\varepsilon_{PA}=10^{-10}$, and the coefficient of the chain rules of smooth min- and max- entropies is set to $\hat{\varepsilon}=10^{-10}$.

We test the performance for all the fiber distances with the parameters optimized for long distance (the ``$+$''-shape points in Fig.~\ref{Fig:KeyRate}). The number of signal pulses sent is $5.18\times10^{14}$ in the 952 km case, $1.02\times10^{14}$ in the 901 km case,  $1.52\times10^{13}$ in the 600 km to 800 km cases, and  $1.26\times10^{12}$ in the 202 km to 499 km cases. The secure key rate in the 952 km case is calculated as $8.75\times10^{-12}$, which is 0.0031 bps. (See Supplemental Materials for details of parameters and results). The secure key rates exceed the absolute PLOB bound~\cite{pirandola2017fundamental} for all the tests with the fiber distances equal to or longer than 398 km, where the PLOB bound is calculated as $-\log_2(1-\eta)$ with the optical and detection efficiency in Charlie set to $\eta_{opt}=100\%$.

Next, we optimized the parameters for short distance, and performed the experiment for 202 km to 398 km again (the ``$\times$''-shape points in Fig.~\ref{Fig:KeyRate}). The ``dim phase reference'' is set to 400 ns in the 1 $\mu s$ signal period, the intensity set to the quantum signal level through all the 100 ms period. The effective signal frequency is thus increased to 900 MHz. The number of signal pulses sent is $3.24\times10^{12}$ in the same collection time as the previous 202 km to 398 km tests. The secure key rate is increased from $3.01\times10^{-5}$ to $5.23\times10^{-5}$ for the 202 km test, with the key rate per second increased from 10.56 kbps to 47.06 kbps.

{\it Conclusion.---}
In conclusion, we experimentally demonstrated an SNS-TF-QKD over 1002 km in infinite regime and up to 952 km considering finite size effect. The main elements enabling the ultra-long distance experiment include the ultra-low-loss fiber, the ultra-low-noise SNSPD, and the time-multiplexed dual-band phase stabilization method. Our experiment demonstrates the feasibility of SNS-TF-QKD through extremely long fiber channel. The secure key rate through shorter fiber is applicable in many practical scenarios. We expect the technology developed in this work will find more general applications in quantum communications~\cite{chen2021integrated}.

\emph{Acknowledgement.}---The authors would like to thank W. Li for the help in developing the phase stabilization method, X.-L. Hu for the help in the secure key rate calculation.

Y.L. and W.-J.Z. contributed equally.

\section*{Supplemental Material}

\section{Theory of SNS-TF-QKD protocol}
In carrying out the SNS protocol~\cite{wang2018sns}, we use 3 sources (intensities): the vacuum source $v$, the decoy source $x$, the signal source $y$ whose intensities are $\mu_v=0,\mu_x,\mu_y$ respectively. In each time window, Alice (Bob) randomly chooses one source $l$ ($r$) from the three candidate sources with probability $p_l$ ($p_r$) for $l=v,x,y(r=v,x,y)$ and then send the prepared pulse to the measurement station which is controlled by Charlie. Charlie is assumed to measure the interference results of the incoming pulse pair and announce the measurement results to Alice and Bob. If Charlie announces one and only one detector clicks, Alice and Bob take the time window as an heralded window, and the corresponding event is an heralded event. After Alice and Bob send $N$ pulse pair to Charlie and receive all the measurement results, they perform the following data post process.

For clarify, we define

$lr$ heralded event or $lr$ heralded window: an event or a time window when Alice sends out a pulse of intensity $\mu_l$ while Bob sends out a pulse of intensity $\mu_r$;

$n_{lr}$ and $S_{lr}$: the number of $lr$ heralded event and the counting rate of $lr$ source where $S_{lr}=\frac{n_{lr}}{np_lp_r}$;

$\mean{M}$: the expected value of the quantity $M$.

For the heralded windows, Alice and Bob first announce the positions of the windows that contains at least one $x$ source, i.e., the time windows that either Alice or Bob chooses the $x$ source. For the rest heralded windows, Alice (Bob) denotes it as bit $0$ ($1$) if she (he) chooses the source $v$ and denotes it as bit $1$ ($0$) if she (he) chooses the source $y$, and those bits form the sifted key string $Z_A$ ($Z_B$) which contains $n_t$ bits. Then Alice and Bob perform the active odd-parity pairing (AOPP)~\cite{xu2020sending} process to decrease the bit-flip error rate:

Bob first randomly pairs the bits $1$ with bits $0$ in $Z_B$ and he obtains $n_g$ odd-parity pairs. Bob announces the positions of those pairs to Alice. Among these $n_g$ pairs, $n_t^\prime$ of them have odd parity at Alice's side and these $n_t^\prime$ pairs will pass the parity check. For the survived pairs, Alice and Bob randomly keep one bit from each pair, and obtain the new sifted key string $Z_A^\prime$ and $Z_B^\prime$ in Alice's side and Bob's side respectively.

Then, Alice and Bob perform the error correction procedure to $Z_A^\prime$ and $Z_B^\prime$ to correct all the wrong bits. After this, Alice shall know all the positions of the wrong bits and she publicly announces them. For the unpaired bits in the random pairing process and un-survived pairs in the parity check process, Alice and Bob publicly announce their bit values.

Given the information above, both of Alice and Bob know the values of $n_{lr}$ for $l,r=v,x,y$. They can use all these values~\cite{hu2022universal} in taking the decoy-state analysis. 

For the time windows that both Alice and Bob choose the sources $x$, Alice and Bob would announce the private phases of the pulse pair and take the following criterion to perform the phase-post selection
\begin{equation}\label{cri}
1-\vert \cos(\theta_{A1}-\theta_{B1}-\psi_{AB})\vert\le \lambda,
\end{equation}
where $\theta_{A1}$ and $\theta_{B1}$ are the private phases of Alice's and Bob's pulses respectively; $\psi_{AB}$ can take an arbitrary value, which can be different from time to time as Alice and Bob like;  $\lambda$ is a small positive value. Denote the number of total pulse pairs sent out that satisfactory Eq.~\eqref{cri} by $N_X$ and the number of corresponding error effective events as $m_X$. Then they (Alice and Bob) can perform the decoy-state analysis to estimate the lower bounds of the untagged bits and the upper bounds of the phase-flip error rates. Finally we have
\begin{equation}\label{keyrate}
R=\frac{1}{N}\{n_1^\prime[1-H(e_1^{\prime \mbox{ph}})]-fn_t^\prime H(E_t^\prime)\}-\gamma^\prime,
\end{equation}
where $R$ is the key rate of per sending-out pulse pair; $n_1^\prime$ is the lower bound of the number of survived untagged bits after AOPP and  $e_1^{\prime \mbox{ph}}$ is the upper bound of the phase-flip error rate of those survived untagged bits after AOPP; $n_t^\prime$ is the number of survived bits after AOPP and $E_t^\prime$ is the corresponding bit-flip error rate in those survived bits; $f$ is the error correction inefficiency and we set $f=1.16$ here; $H(x)=-x\log_2x-(1-x)\log_2(1-x)$ is the Shannon entropy. The parameter values such as $n_1'$ and $e_1^{'ph}$ can be verified by decoy-state analysis. As shall be shown latter, the additional term $\gamma^\prime$ takes logarithm small values and it is for the security with finite-data size and the advanced decoy state analysis. 

We have the lower bounds of the expected values of the counting rate of states $\oprod{01}{01}$ and $\oprod{10}{10}$ below

\begin{align}
\label{s01mean}\mean{\underline{s_{01}}}&= \frac{\mu_{y}^{2}e^{\mu_{x}}\mean{S_{vx}}-\mu_{x}^{2}e^{\mu_{y}}\mean{S_{vy}}-(\mu_{y}^{2}-\mu_{x}^{2})\mean{S_{vv}}}{\mu_{y}\mu_{x}(\mu_{y}-\mu_{x})},\\
\mean{\underline{s_{10}}}&= \frac{\mu_{y}^{2}e^{\mu_{x}}\mean{S_{xv}}-\mu_{x}^{2}e^{\mu_{y}}\mean{S_{yv}}-(\mu_{y}^{2}-\mu_{x}^{2})\mean{S_{vv}}}{\mu_{y}\mu_{x}(\mu_{y}-\mu_{x})}.
\end{align}
Then we can get the lower bound of the expected value of the counting rate of untagged photons
\begin{equation}
\mean{\underline{s_1}}=\frac{1}{2}(\mean{\underline{s_{10}}}+\mean{\underline{s_{01}}}),
\end{equation}
and the lower bound of the expected value of the untagged bits $1$, $\mean{\underline{n_{10}}}$, and untagged bits $0$, $\mean{\underline{n_{01}}}$
\begin{align}
\mean{\underline{n_{10}}}=Np_vp_{y}\mu_{y}e^{-\mu_{y}}\mean{\underline{s_{10}}},\\
\mean{\underline{n_{01}}}=Np_vp_{y}\mu_{y}e^{-\mu_{y}}\mean{\underline{s_{01}}}.
\end{align}

And the upper bound of the expected value of the phase-flip error rate satisfies
\begin{equation}\label{e1}
\mean{\overline{e_1^{ph}}}=\frac{\mean{T_{X}}-e^{-2\mu_x}\mean{S_{vv}}/2}{2\mu_xe^{-2\mu_x}\mean{\underline{s_1}}},
\end{equation}
where $\mean{T_{X}}$ is the expected value of $T_{X}$, and $T_X=m_X/N_X$.

For the asymptotic case, we can directly regard $\mean{M}=M$ and $\gamma^\prime=0$. And we can estimate the lower bound of the number of survived untagged bits and the upper bound of the phase-flip error rate after AOPP by the following formulas:
\begin{align}
&n_1^\prime=\frac{\mean{\underline{n_{10}}}}{n_{t1}}\frac{\mean{\underline{n_{01}}}}{n_{t0}}n_g,\\
&e_1^{\prime \mbox{ph}}=2\mean{\overline{e_1^{ph}}}(1-\mean{\overline{e_1^{ph}}}),
\end{align}
where $n_{t0}$ and $n_{t1}$ are the number of bits $0$ and bits $1$ in the raw key string $Z_B$ respectively. With those two values, the key rate in the asymptotic case can be calculated by applying Eq.~\eqref{keyrate}.

For the non-asymptotic case, the expected values can be estimated according to their observed values by applying Chernoff bound. We have the related formulas of $n_1^\prime$ as follows~\cite{xu2020sending}:
\begin{subequations}
\begin{align}
&u=\frac{n_g}{2n_{odd}},\\
&\underline{n_{10}}=\varphi^L(u\mean{\underline{n_{10}}}),\\
&\underline{n_{01}}=\varphi^L(u\mean{\underline{n_{01}}}),\\
&\underline{n_{1}}=\underline{n_{10}}+\underline{n_{01}},\\
&n_1^r=\varphi^L\left(\frac{\underline{n_{1}}^2}{2un_t}\right),\\
&n_{01}^\prime=2n_1^r\left(\frac{\underline{n_{01}}}{\underline{n_{1}}}-\sqrt{-\frac{\ln\varepsilon}{2n_1^r}}\right)\\
&n_{10}^\prime=2n_1^r\left(\frac{\underline{n_{10}}}{\underline{n_{1}}}-\sqrt{-\frac{\ln\varepsilon}{2n_1^r}}\right)\\
&n_{min}=\min(n_{01}^\prime,n_{10}^\prime),\\
&n_1^\prime=2\varphi^L\left(n_{min}(1-\frac{n_{min}}{2n_1^r})\right),\\
\end{align}
\end{subequations}
where $n_t$ is number of raw keys that Alice and Bob get in the experiment; ${n_{odd}}$ is the number of pairs with odd-parity if Bob randomly groups all his raw key bits two by two, and $n_g$ and $n_{odd}$ are observed values; $\epsilon$ is the failure probability of parameter estimation; and $\varphi^U(x),\varphi^L(x)$ are the upper and lower bounds while using Chernoff bound~\cite{chernoff1952measure} to estimate the real values according to the expected values.

And we have the the related formulas of $e_{1}^{\prime ph}$ as follows:
\begin{subequations}
\begin{align}
&r=\frac{\underline{n_1}}{\underline{n_1}-2n_1^r}\ln\frac{3(\underline{n_1}-2n_1^r)^2}{\varepsilon},\\
&e_{\tau}=\frac{\varphi^U(2n_1^r\mean{\overline{e_1^{ph}}})}{2n_1^r-r}\\
&\bar{M}_s=\varphi^U[(n_1^r-r){e_{\tau}}(1-{e_{\tau}})]+r,\\
&e_1^{\prime \mbox{ph}}=\frac{2\bar{M}_s}{n_1^\prime}
\end{align}
\end{subequations}

Finally, the key rate in the non-asymptotic case can be calculated by applying Eq.~\eqref{keyrate} and the extra term $\gamma^\prime$ is~\cite{hu2022universal}
\begin{equation}\label{r2}
\begin{split}
\gamma^\prime=\frac{1}{N}[2\log_2{\frac{2}{\varepsilon_{cor}}}+4\log_2{\frac{1}{\sqrt{2}\varepsilon_{PA}\hat{\varepsilon}}}\\+2\log_2(n_t-n_{vv}-n_{yy})],
\end{split}
\end{equation}
where $\varepsilon_{cor}$ is the failure probability of error correction, $\varepsilon_{PA}$ is the failure probability of privacy amplification, $\hat{\varepsilon}$ is the coefficient while using the chain rules of smooth min- and max- entropy~\cite{vitanov2013chain}, and $2\log_2(n_t-n_{vv}-n_{yy})$ is the extra cost of the advanced decoy state analysis~\cite{hu2022universal}. 

Compared with the 3-intensity SNS protocol proposed in Ref.~\cite{yu2019sending}, there is no need to set a separate vacuum source or non-vacuum source of intensity $\mu_y$ for decoy state analysis in the advanced 3-intensity SNS protocol~\cite{hu2022universal}. Such an improvement, on the one hand, can increase the number of counts related to the vacuum source and $\mu_y$ source in the decoy-state analysis and hence reduce the statistical fluctuation effect; on the other hand, can increase the number of raw keys, and finally improve the key rate.

\section{Dual-band stabilization method with data post-processing}
In our experiment, we developed a dual-band stabilization method with data post-processing for estimating the phase between Alice's and Bob's channels. The method do not need additional active phase modulators inserted, so the total loss in Charlie is smaller. The method do not need real-time feedback circuits, so it is more flexible in both algorithm design and the required hardware resource.

Similar to hard-ware based dual-band stabilization method, two wavelengths are used in our method. The $\lambda_1$ ``strong phase reference'' is wavelength multiplexed with the quantum signal. This $\lambda_1$ light is used for estimating the fast phase fluctuation in the channel. The $\lambda_2$ ``dim phase reference'' is time multiplexed with the quantum signal. This $\lambda_2$ light is used for correct residual phase after ``strong phase reference''.
In the experiment, the $\lambda_1$ light is modulated to a 400 ns pulse in 1 $\mu$s period, to avoid inducing noises on the quantum signal. In this 400 ns pulse period, Alice and Bob modulate the phase to $\phi_A = \{0,0,\pi,\pi\}$, and $\phi_B = \{0,\pi/2,\pi/2, 0\}$, each in a 100 ns span. This will yield a phase difference between Alice and Bob of $\delta_{AB}=\phi_{B}-\phi_{A} = \{0,\pi/2,-\pi/2, \pi\}$. The sequence will be recored and adjusted in the data post-processing. In this design, in either Alice or Bob, only two different phases are required, and the highest voltage is $V_\pi$ instead of $V_{3\pi/2}$ in direct generation four phases. This will lower the requirement of the signal generators, amplifiers, and modulators.

The $\lambda_2$ light is first modulated to a 1 GHz pulse train with a width of 120 ps, as the designed pulse width for the quantum signals. The pulse train is then modulated with a 100 ms period. The intensity of the pulse train in the first 40 ms is set to a higher level as the ``dim phase reference''; and single photon levels in the rest 60 ms as the quantum signal. In the longest distance for example, the detection rate of ``dim phase reference'' expected to be about 1000 Hz at Charlie. This empirical parameter is selected to guarantee the accurate estimation of the initial phase difference between $\lambda_1$ and $\lambda_2$. Then, the $\lambda_2$ pulses are encoded in a 1 $\mu$s period synchronized with the ``strong phase reference'', where the first 400 ns are used for the ``dim phase reference'' and the rest 600 ns are used for the quantum signal.
The encoding of the 400 ns ``dim phase reference'' is similar to the modulations in quantum signal: 400 sets of random numbers are used to modulate the intensities and phases of the pulses. For each ``dim phase reference'' pulse, the intensity is set to one of $\{0, \nu, \mu\}$; the phase is set to one of the 4 phases of $\{0,\pi/2,-\pi/2, \pi\}$. Different from the quantum signal modulation, the random numbers used for ``dim phase reference'' are correlated and public announced . The intensities of the pulses of this 400 ns period are the same for Alice and Bob, the phases are selected that the difference between Alice's and Bob's modulation are $\delta_{AB} = \{0,\pi/2,-\pi/2, \pi\}$ for each 100 ns span.

The phase reference pulses are then transmitted and interfered at Charlie. The detection rate of the ``strong phase reference'' is set in the range between 0.3 MHz and 1.5 MHz for each detector, based on the specific experimental condition. The detection of ``dim phase reference'' is set to about 1 kHz detections for each detector in the longest fiber distance, and set to a higher level for shorter distances.

As discussed in main text, we consider the effect of wavelength difference, and estimate the relative phase induced in fiber spools of the quantum signal with:
\begin{equation}
	\phi_s(t) = \phi_r(t) \times \lambda_2/\lambda_1 + \phi_{sr}(0)
	\label{Eq:FCorrectionSM}
\end{equation}
where $\phi_r(t)$ is the phase difference of the ``strong phase reference'', and $\phi_{sr}(0)=\phi_s(0)-\phi_r(0)$ is the initial phase difference between $\lambda_1$ and $\lambda_2$.

We have detections for four known phase differences of ``strong phase reference'' signals and ``dim phase reference'' signals, based on Alice's and Bob's modulation pattern. The question is how to calculate the phase difference induced by the channel, from the ``strong phase reference''; and how to calculate the phase difference of the first pulse, to finally calculate the phase difference of each quantum signal.

Assume the phase of a ``strong phase reference'' signal modulated by Alice (Bob) is $\theta_A$ ($\theta_B$), the phase induced by the fiber spools of Alice-Charlie (Bob-Charlie) is $\varphi_A$ ($\varphi_B$), then the phase difference between Alice's and Bob's modulation is $\Delta\theta=\theta_A-\theta_B$, the phase difference between the fiber links is $\Delta\varphi_T=\varphi_A-\varphi_B$. The phase difference at Charlie is:
\begin{equation}
\phi=\theta_A-\theta_B+\Delta\varphi_T
\end{equation}

The normalized intensity of Charlie's interference output will be:
\begin{equation}
\begin{aligned}
I(\phi)&=\vert1+\mathrm{e}^{\mathrm{i}\phi}\vert^2/2=[1+\cos(\phi)]/2=\cos^2(\phi/2)
\label{Eq:Interference}
\end{aligned}
\end{equation}
For the single photon detections, the normalized intensity at Charlie stands for the probabilities Charlie detects a signal from the detector at the first output port of the BS.

For the $\lambda_1$ ``strong phase reference'', we take a statistic period of 40 $\mu$s, summing the detections of each phase difference $\Delta\theta$ as $N_0$, $N_{\pi/2}$, $N_{\pi}$ and $N_{3\pi/2}$. Then we calculate the probabilities of each phase difference, as:
\begin{equation}
p_{i}=2 N_{i}/\Sigma N_{i}
\label{Eq:Prop}
\end{equation}
where $i=1...4$ stands for the scenario the phase difference between Alice and Bob are
\{$0$, $\pi/2$, $\pi$, $3\pi/2$\}.

We build a error model based on Eq.(\ref{Eq:Interference}):
\begin{equation}
Err(\Delta\varphi) = \sum_i{p_i\cdot(1-\cos((\Delta\theta_i+\Delta\varphi)/2)^2)^2}
\label{Eq:Err1}
\end{equation}
where $\Delta\theta_i=$\{$0$, $\pi/2$, $\pi$, $3\pi/2$\} are the nominal phase differences that between Alice's and Bob's modulations, $p_i$ is the detected probabilities of each phase difference, $\Delta\varphi$ is the guessed phase difference between Alice's and Bob's fiber. Note that the probability $p_i$ is determined by both the phase difference $\Delta\theta_i$ and the phase difference of the fiber. The result of the error model of Eq.(\ref{Eq:Err1}) will reach its minimum when the guessed phase difference $\Delta\varphi$ is equal to the actual phase difference of the fiber. We traverse $\Delta\varphi$ from 0$^\circ$ to 359$^\circ$ with 1$^\circ$ as a step, to search for the minimum $Err(\Delta\varphi)$, and thus the most possible phase difference $\Delta\varphi$ induced by the channel. In the experiment, the phase difference of the ``strong phase reference'' signal $\phi_r(t)$ is set to this estimated phase difference $\Delta\varphi$.

In experiment, we calculate the channel phase for $\lambda_1$ ``strong phase reference'' for each 40 $\mu$s. Due to the definition domain of the trigonometric functions, the calculated phase falls in the range of [0, 2$\pi$). We assume the drift of the channel phase is relatively slow. In particular, we assume the  phase differences induced by fiber channel between the adjacent 40 $\mu$s calculation is smaller than 180$^\circ$. So we can resort the relative phase between the fiber channel to a continuous form, following the least step assumption. After this modification, the $\phi_r(t)$ reflect the actual phase drift of the fiber channel. Only with this range expansion, the relative phase between different wavelengths can be calculated more accurate using Eq.(\ref{Eq:FCorrectionSM}).

Now the only parameter left is the initial phase difference $\phi_s(0)-\phi_r(0)$ between the wavelengths. We calculate this initial phase difference based on both the ``strong phase reference'' and ``dim phase reference'' signal. Recall that for each ``dim phase reference'' detection $i$ with a phase difference $\Delta\theta_i$, we have calculated a channel phase difference $\Delta\varphi_i$ based on the ``strong phase reference''. The total phase difference  for $\lambda_2$ is:
\begin{equation}
\Delta\varphi_{s,i} = \Delta\varphi_i\times\lambda_2/\lambda_1+\Delta\theta_i+\phi_{sr}(0)
\end{equation}
where $\phi_{sr}(0)=\phi_s(0)-\phi_r(0)$ is the initial phase difference between $\lambda_1$ and $\lambda_2$, that we need to calculate.

Next we discretize the phase $\Delta\phi_{s,i}=\Delta\varphi_i\times\lambda_2/\lambda_1$ to a precision of 1$^\circ$. We count the number of ``dim phase reference'' detections as $n_i$ with different phases $\Delta\phi_{s}$. Now we have a array of $n_i$ with $i=0...359$, standing for the number of detections of ``dim phase reference'', with $\Delta\phi_{s}$ falls in the corresponding phase bin.

We use a error model similar to Eq.(\ref{Eq:Err1}):
\begin{equation}
Err(\Delta\varphi') = \sum_i{n_i\cdot(1-\cos((\Delta\theta_i+\Delta\varphi')/2)^2)^2}
\label{Eq:Err2}
\end{equation}
The result of the error model will reach its minimum when the guessed initial phase difference $\Delta\varphi'$ is equal to the actual initial phase difference. We traverse $\Delta\varphi'$ from 0$^\circ$ to 359$^\circ$, with 1$^\circ$ as the step, and search for the minimum $Err(\Delta\varphi')$. Then the initial phase difference $\phi_{sr}(0)$ between the quantum signal and the ``strong phase reference''  is set to this estimated phase difference $\Delta\varphi'$.
We note that although the error model Eq.(\ref{Eq:Err2}) is similar to Eq.(\ref{Eq:Err1}), there are two differences. The first is that the number of detections is used instead of probability in implementation. The second is that 360 possible phases are used instead of 4 phase differences, since the phase difference of $\lambda_1$ in the equation changes continuously. 

Having the initial phase difference $\phi_{sr}(0)$, and the estimated channel phase difference $\Delta\varphi_{r,i}$ obtained with the ``strong phase reference'', the channel phase difference $\Delta\varphi_{s,i}$ of the quantum signals can be calculated with Eq.(\ref{Eq:FCorrectionSM}), for each quantum signal detection.

\section{Measurement of noises}

\begin{figure}[tbh]
\centering
\resizebox{8.5cm}{!}
{\includegraphics{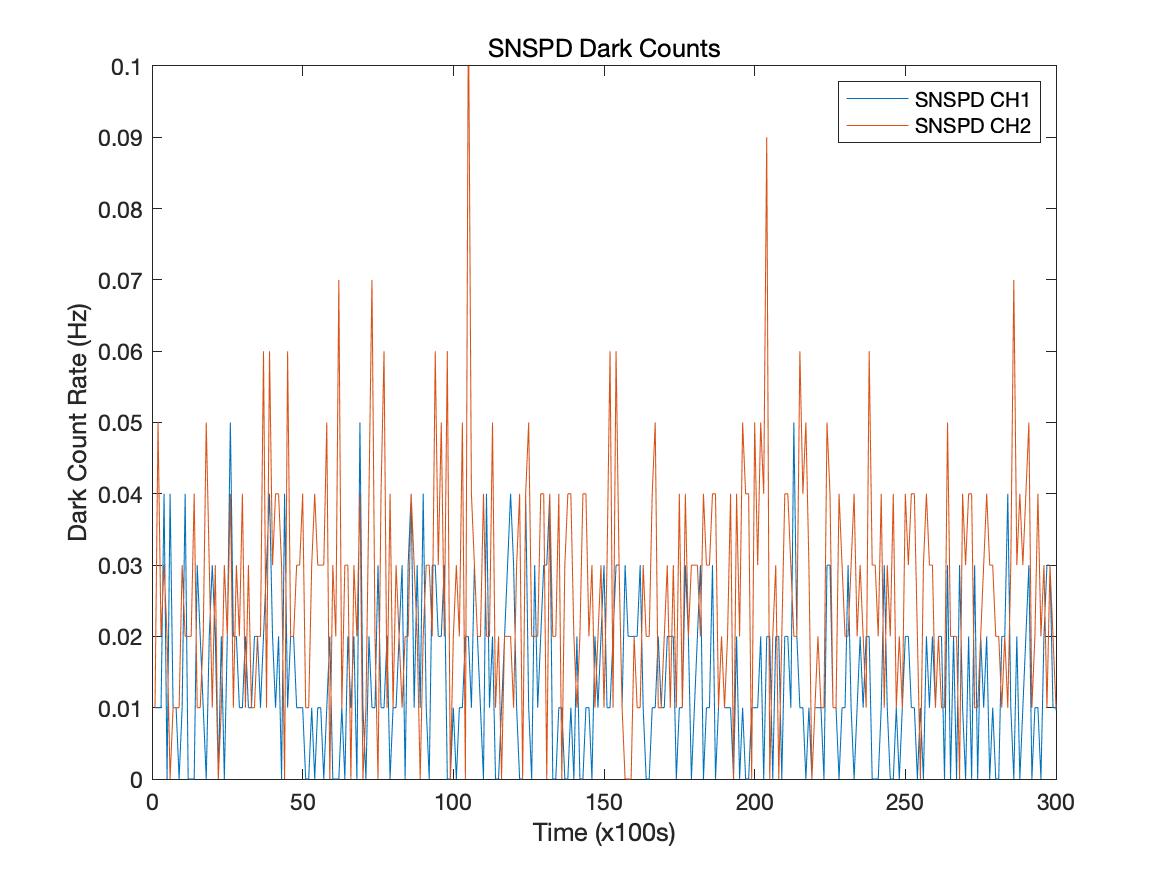}}
\caption{Dark count rate of the ultra-low noise SNSPDs. The average dark count rates of the two channels are 0.014 Hz and 0.026 Hz.}
\label{Fig:SM:DCR}
\end{figure}

{\it SNSPD dark counts.---}
We measured the superconducting nanowire single-photon detector (SNSPD) dark count. The average dark count rate of the two channels are 0.014 Hz and 0.026 Hz on installation, as shown in Fig.~\ref{Fig:SM:DCR}. We note that the dark count rate increases during our experiment. 

{\it Noise measurement using strong reference.---}

\begin{figure}[tbh]
\centering
\resizebox{8.5cm}{!}
{\includegraphics{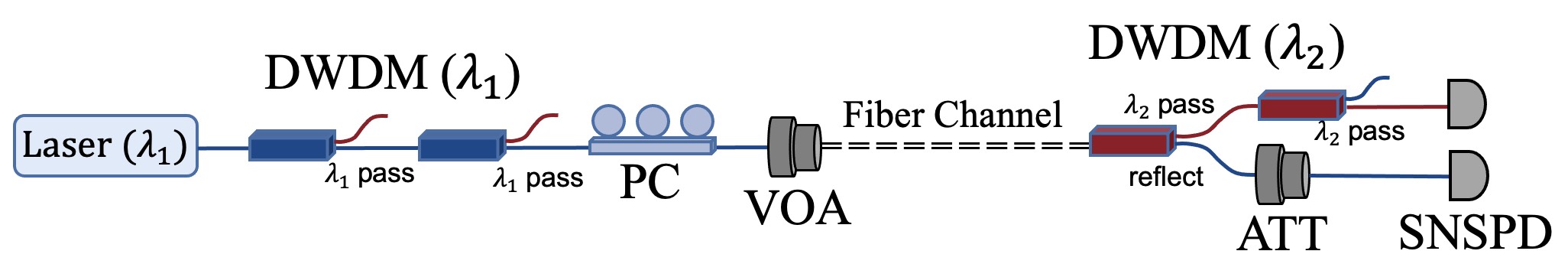}}
\caption{Noise Measurement setup. DWDM, dense wavelength division multiplexing; PC, polarization controller; VOA, variable optical attenuator; ATT, attenuator; SNSPD, superconducting nanowire single-photon detector.}
\label{Fig:SM:NNSetupAtt}
\end{figure}

We measure the noises induced by the ``strong phase reference'' light with two methods. The first method is to use stronger reference light and attenuators for accurate measurement. The setup is shown in Fig.~\ref{Fig:SM:NNSetupAtt}, the laser is set to $\lambda_1$ the ``strong phase reference'' wavelength, following two DWDMs for filtering. A variable optical attenuator (VOA) is used to set the intensity before the fiber channel. At the receiver, DWDMs are used to split and filter the light, for the measurement of noise induced in $\lambda_2=1550.12$ nm, and the intensity of the $\lambda_1$ ``strong phase reference''. In this test, we used an SNSPD with a dark count rate of about 0.73 Hz.

\begin{figure}[tbh]
\centering
\resizebox{8.5cm}{!}
{\includegraphics{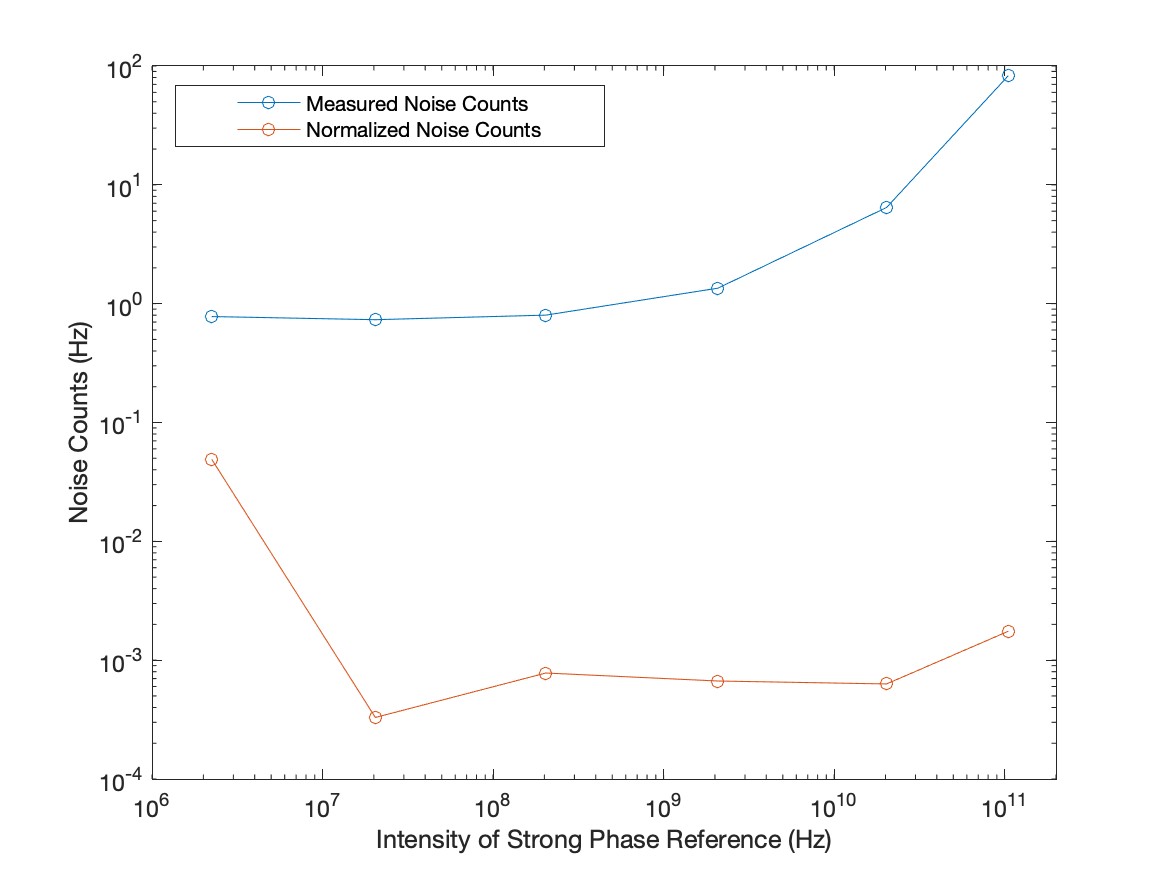}}
\caption{Measured noises with 0 km fiber. The blue circles shows the measured noises. The orange circles shows the noises subtracting the SNSPD noise and normalized to 2 MHz ``strong phase reference''. }
\label{Fig:SM:NN0km}
\end{figure}

First, we set the distance of the fiber channel to 0 km, and inserted a 40 dB attenuator in the ``strong phase reference'' path after the DWDMs and before the SNSPD. We set the detections of the ``strong phase reference'' channel to a range of \{200 Hz,... 10 MHz\}. Considering the 40 dB attenuator, the ``strong phase reference'' intensity $N_{ref}$ is actually around \{2 MHz,... 100 GHz\}. The original measurement result of the noise counts $N_{raw}$ is shown in Fig.~\ref{Fig:SM:NN0km}. The noise approaches the SNSPD dark count rate when the ``strong phase reference'' intensity $N_{ref}$ is below $2\times10^8$. We assume the induced noise scales linearly with the ``strong phase reference'' detections, then we subtract the dark counts and normalize the noise to a 2 MHz ``strong phase reference'' detections:
\begin{equation}
	N_{noise} = (N_{raw}-N_{dcr})\times2\times10^6/N_{ref}
\end{equation}
where $N_{dcr}\approx 0.73$ Hz is the SNSPD dark count rate. The result shows a cross talk noise of around $10^{-3}$. The noise tested with a higher ``strong phase reference'' intensity $N_{ref}$ is more accurate for the fluctuation is smaller.

\begin{figure}[tbh]
\centering
\resizebox{8.5cm}{!}
{\includegraphics{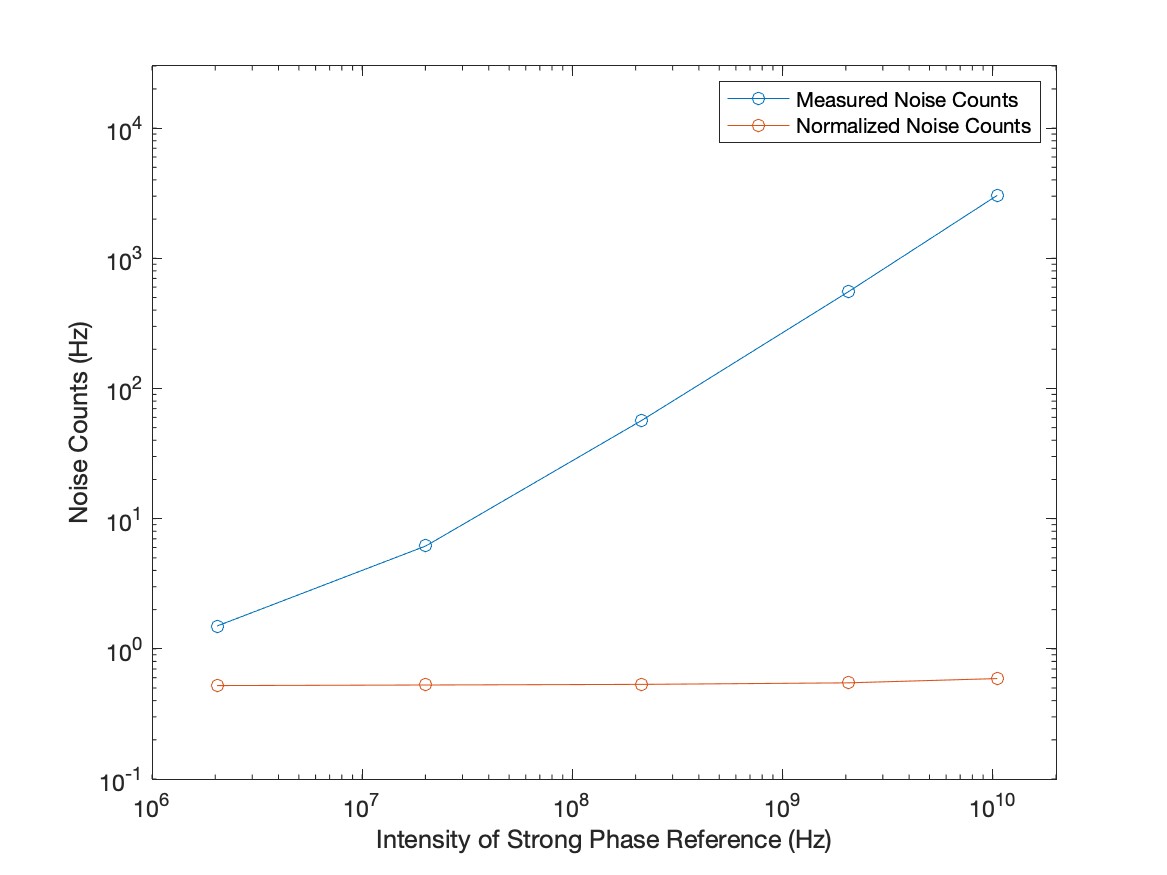}}
\caption{Measured noises with 300 km fiber. The blue circles shows the measured noises. The orange circles shows the noises subtracting the SNSPD noise, and normalized to 2 MHz ``strong phase reference''.}
\label{Fig:SM:NN300km}
\end{figure}

Next, we set the distance of the fiber channel to 300 km, and measure the noise with the same procedure. The noise is $N_{noise}=0.55$ Hz when subtracting the dark counts and normalizing the ``strong phase reference'' to 2 MHz.

\begin{figure}[tbh]
\centering
\resizebox{8.5cm}{!}
{\includegraphics{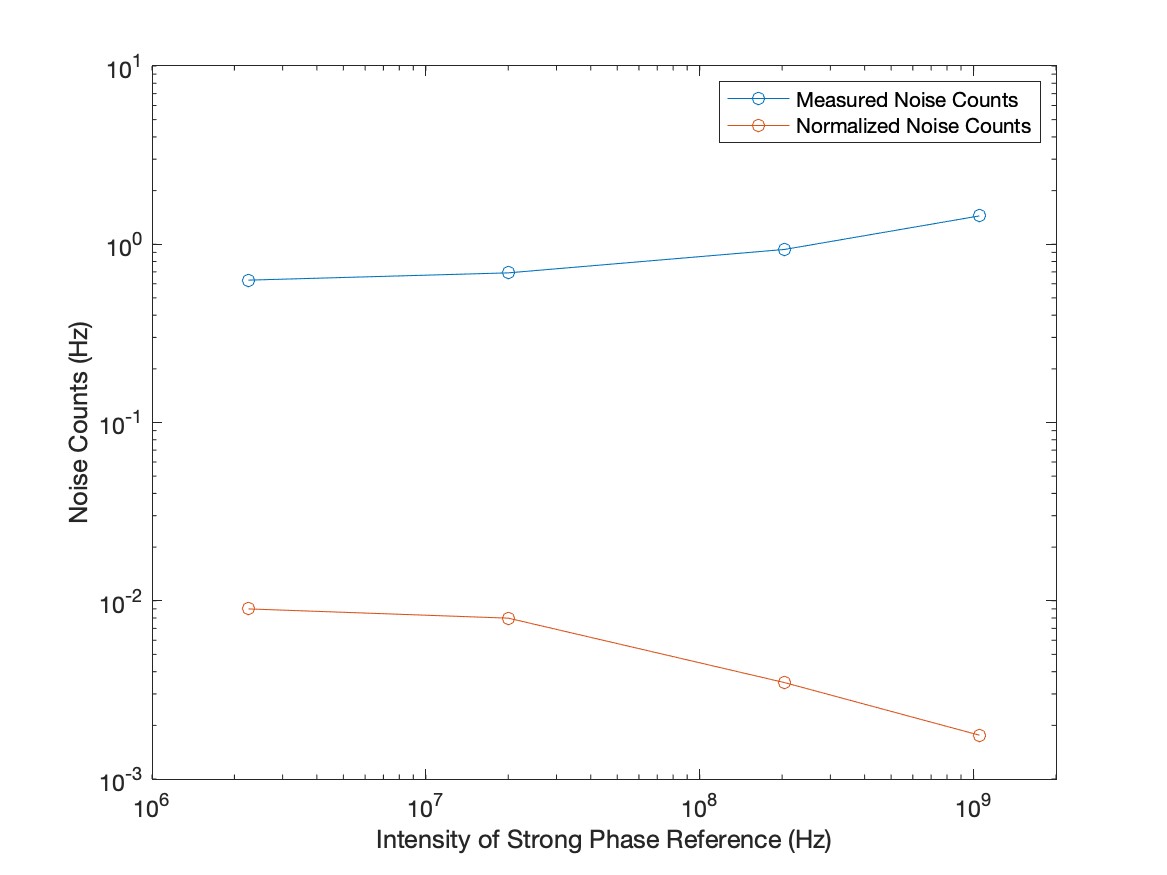}}
\caption{Measured noises with 300 km fiber when ``strong phase reference'' is modulated to 1 $\mu$s pulse in a 2 $\mu$s period. The data is collected in the last 1 $\mu$s when the pulse is turned off. The blue circles shows the measured noises. The orange circles shows the noises subtracting the SNSPD noise, and normalized to 2 MHz ``strong phase reference''.}
\label{Fig:SM:NN300kmP}
\end{figure}

Finally, we test the noise with the pulsed strong light through 300 km fiber. The ``strong phase reference'' is modulated to a 1 $\mu$s pulse in a 2 $\mu$s period. This time, we count the noise counts in the last 1 $\mu$s when the ``strong phase reference'' turned off. The measured noise is $N_{noise}<0.01 $ Hz after subtracting the dark counts and normalize the ``strong phase reference'' to 2 MHz.

{\it Noise measurement using the experimental setup.---}
\begin{figure}[tbh]
\centering
\resizebox{8.5cm}{!}
{\includegraphics{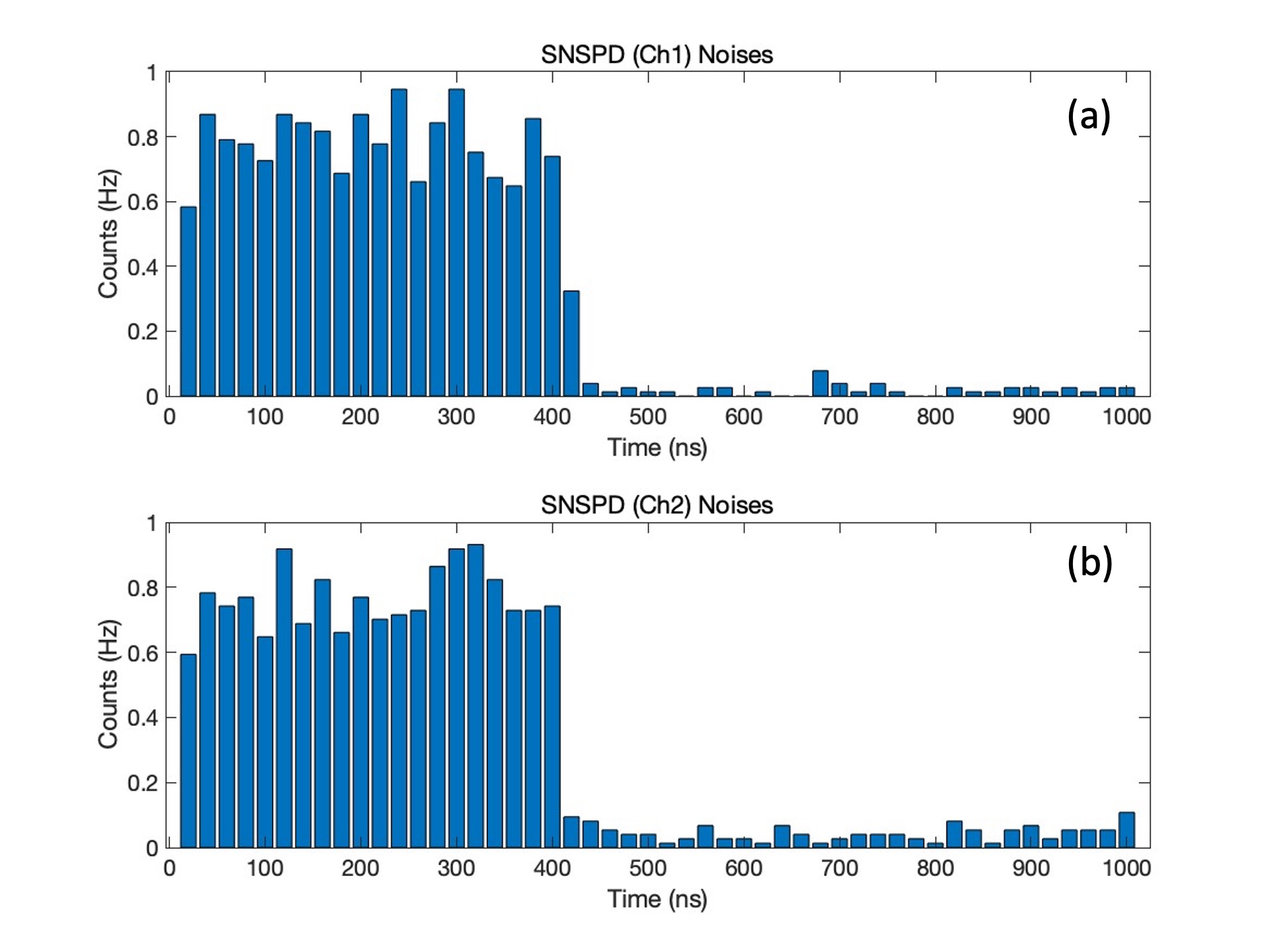}}
\caption{Measured noises with 500 km fiber, when the ``strong phase reference'' is modulated to a 400 ns pulse in a 1 $\mu$s period, and the intensity is set to about 1 MHz. The average noise is (a) 0.8 Hz and (b) 0.75 Hz in the first 400 ns, and (a) 0.019 Hz and (b) 0.035 Hz in the last 600 ns}
\label{Fig:SM:NN500km}
\end{figure}

The second method is to measure the noise induced by fiber directly, using the ultra low dark count SNSPD. This time, we adopted the final experiment setup, as shown in the main text. The intensity of the ``strong phase reference'' is set to about 1 MHz, similar to that used in the experiment.

The measured noises of the $\lambda_2$ quantum channel are shown in Fig.~\ref{Fig:SM:NN500km}. The average noise is about 0.8 Hz and 0.75 Hz in the first 400 ns, where the ``strong phase reference'' is on, and about 0.019 Hz and 0.035 Hz in the last 600 ns, when the ``strong phase reference'' is off. Compared with the SNSPD dark counts of 0.014 Hz and 0.026 Hz, the noise rate is negligible when the ``strong phase reference'' is off.

This result is consistent with the previous test with strong reference light. The noise is small when the laser is connected directly to the receiver, and is increased when a long distance fiber is connected. When the ``strong phase reference'' is modulated to a pulse, the noise is high in the time span the ``strong phase reference'' is on, but decrease to a small value short after the pulse is off. We attribute the main source of noises in the first 400 ns period to the spontaneous Raman scattering from the fiber, and the dominates noise in the rest 600 ns to the SNSPD dark counts.

\section{Real-Time feedback system}
\subsection{Signal Arrival Time Feedback}

\begin{figure}[tbh]
\centering
\resizebox{8.5cm}{!}
{\includegraphics{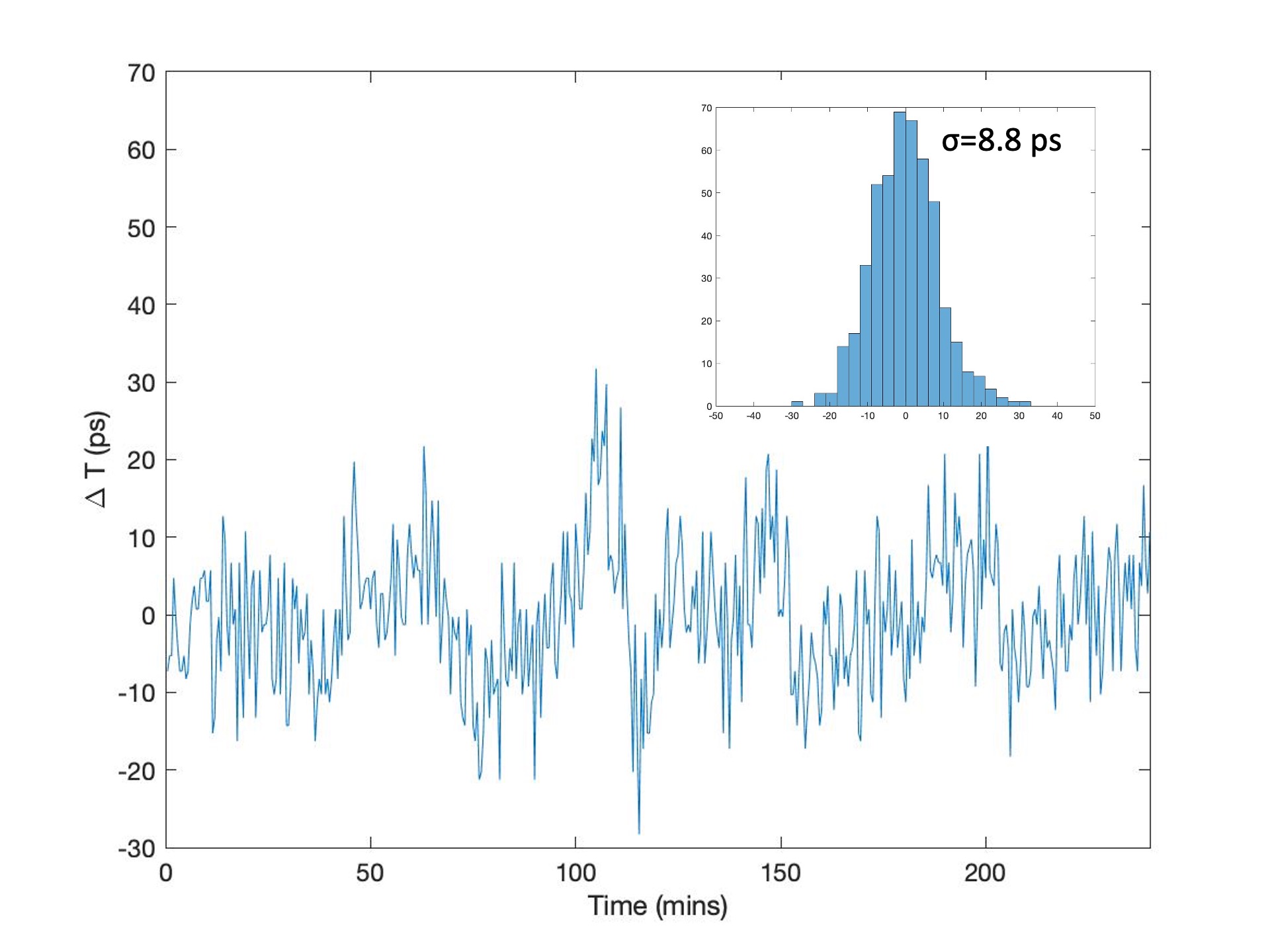}}
\caption{Measured time difference between $\lambda_1$ and $\lambda_2$ in 4 hours. The standard deviation of the time difference is 8.8 ps.}
\label{Fig:SM:DeltaT}
\end{figure}

In the experiment, Alice's and Bob's quantum signals should arrival at the same time for the best interference. Here we used the arrival time of the $\lambda_1$=1548.51 nm ``strong phase reference'' light as an indicator to adjust the arrival time of the quantum signal. 

Before working on the details of the feedback method, we first prove the possibility of the method. We modulate a sequence of pulses of the two wavelengths by modulating the combined light in an intensity modulator. The rising edges of the $\lambda_1$ and $\lambda_2$ pulse begin at the same time. Then the light is sent through the channel. The arrival time through long-distance fiber may drift a few nano-seconds during one day. We record the arrival time of the two wavelengths, using the rising edge method mentioned below. The result is shown in Fig.~\ref{Fig:SM:DeltaT}. The time difference of the rising edges of the two wavelengths is stable, showing a less than 10 ps standard deviation during the test. As a short conclusion, the error induced in estimating the $\lambda_2$ arrival time with $\lambda_1$ is acceptable. 

\begin{figure}[tbh]
\centering
\resizebox{7.5cm}{!}
{\includegraphics{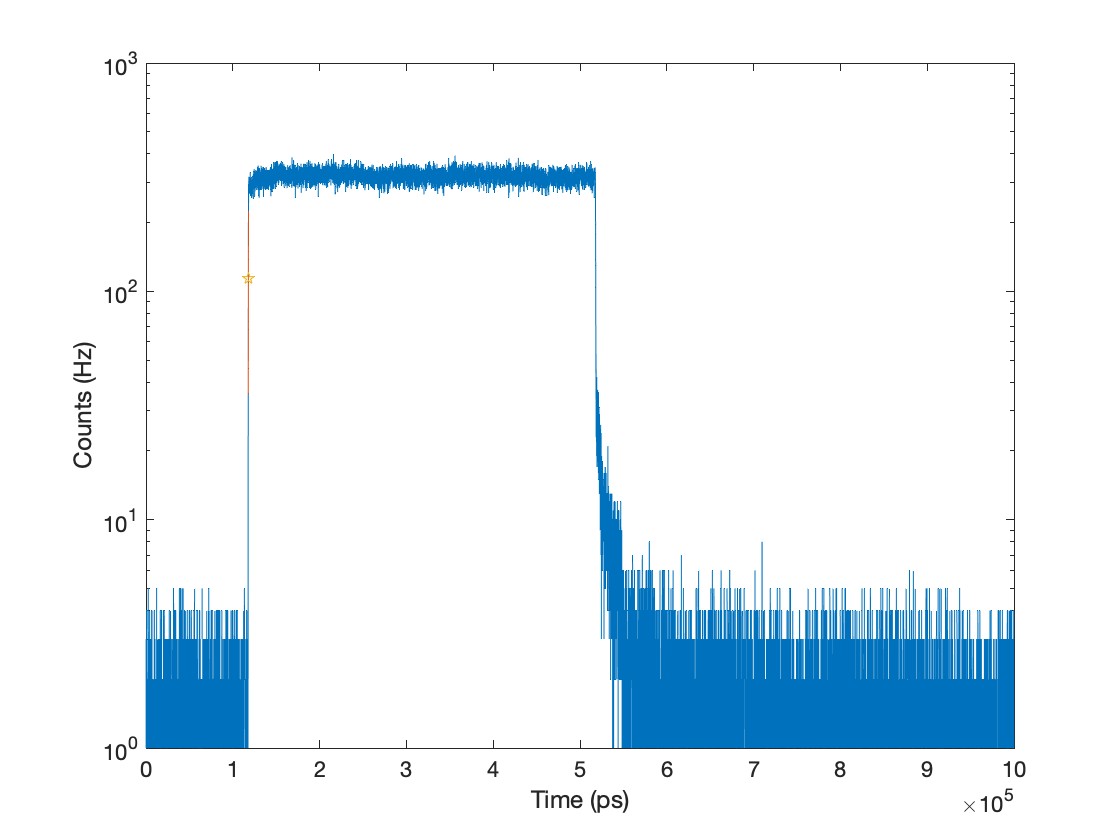}}
\caption{Measured $\lambda_1$ ``strong phase reference'' pulse at Charlie. The orange line is the fitted rising edge of the pulse, the star is the calculated arrival time.}
\label{Fig:SM:RefPulse}
\end{figure}

To acquire the arrival time of $\lambda_1$ in the final experiment setup, the $\lambda_1$ light is separated with a DWDM connected to the monitor port of the input PBS at Charlie. This $\lambda_1$ light is used for both delay feedback, and polarization control. 

Following the modulation sequence, 1 MHz $\lambda_1$ pulses with 400 ns width are sent every second. 

We accumulate the histogram of the $\lambda_1$ pulse with a 10 ps precision, as shown in Fig.~\ref{Fig:SM:RefPulse}. Then we fit the rising edge of the pulse with a linear fitting, using the data between 10\% and 60\% of the peak count. The pulse peak count is calculated using the average count within [20 ns, 220 ns] range after the rising edge. The arrival time is set as the time of the fitting exceed 35\% of the peak count.

\begin{figure}[tbh]
\centering
\resizebox{8.5cm}{!}
{\includegraphics{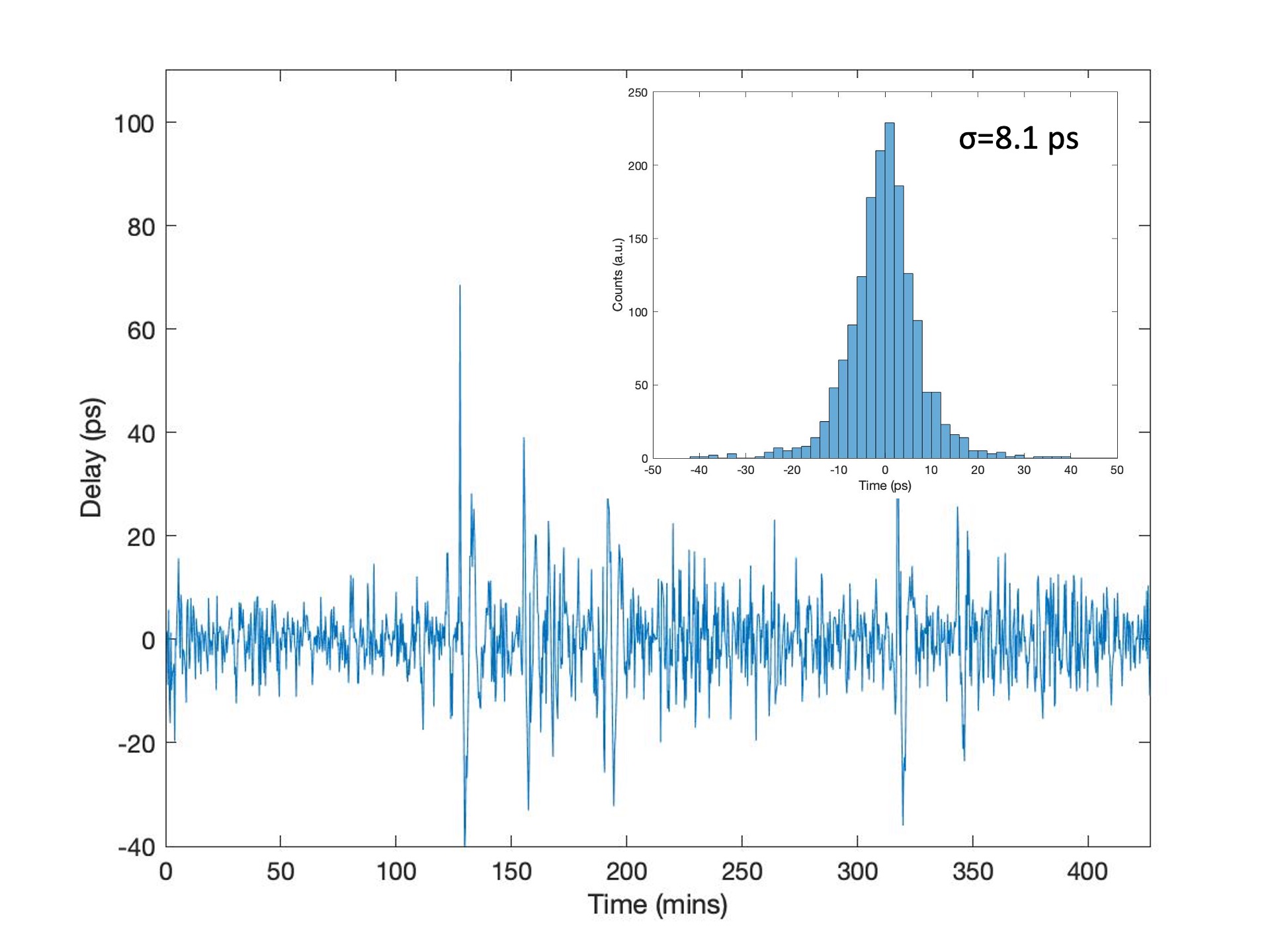}}
\caption{Measured relative delay with the delay feedback on. The standard deviation of the signal arrival time is 8.1 ps.}
\label{Fig:SM:DelayStablized}
\end{figure}

\begin{figure}[tbh]
\centering
\resizebox{8.5cm}{!}
{\includegraphics{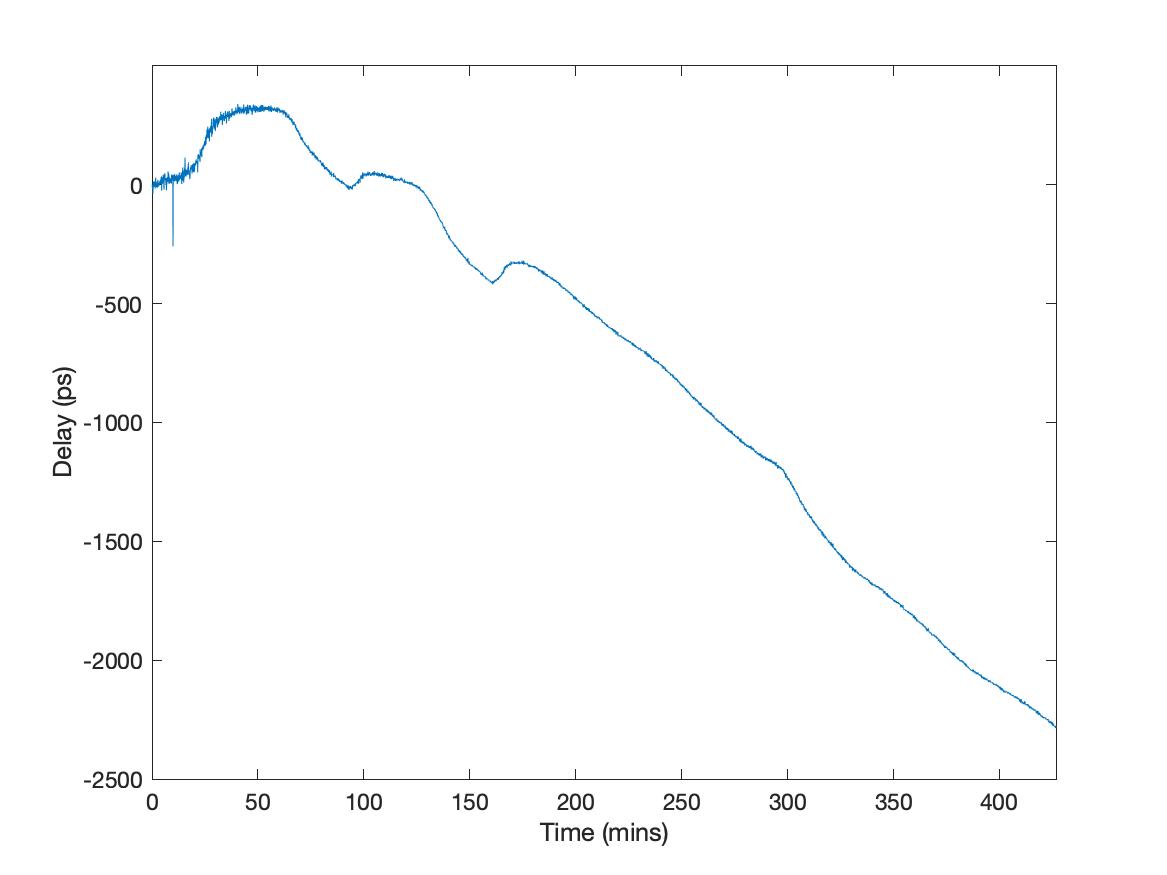}}
\caption{Measured relative delay drift with the delay feedback off.}
\label{Fig:SM:DelayDrift}
\end{figure}

There are fluctuations in calculating the arrival time, due to the fluctuation of the signals, the limited detection counts and the estimation method. To achieve a stable time feedback, we employ a proportional integral derivative (PID) algorithm. The feedback output controls the central clock for Alice and Bob, to stabilize the relative delay. The measured arrival time with and without the delay feedback are shown in Fig.~\ref{Fig:SM:DelayStablized} and Fig.~\ref{Fig:SM:DelayDrift}. The arrival time drifts 2 ns in 7 hours when feedback is off, while it is stabilized to a standard deviation of 8.1 ps with the feedback on.

\subsection{Two Wavelength Polarization Feedback}
In the experiment, we used two different wavelengths, $\lambda_1$=1548.51 nm and $\lambda_2$=1550.12 nm, for ``strong phase reference'' and ``dim phase reference''. We realize the polarization compensation by sending the control sequence to Alice (Bob) based on the measurement result from the PBS monitors port at Charlie. Alice (Bob) then adjust the polarization by controlling a 4-paddles electronic polarization controllers.

As mentioned above, the $\lambda_1$ light in the feedback channel is used for the relative delay feedback, besides its main function of ``strong phase reference''. So the count rate of the $\lambda_1$ light should not be too small; the count rate should not be too high, to accumulate enough ``strong phase reference'' interfere counts for phase estimation. We expect 10\% to 20\% of $\lambda_1$ light to be measured at the monitor port of the PBS for delay feedback, while the majority of the $\lambda_1$ light should be used for phase estimation. We expect the $\lambda_2$ light count rate at the monitor port as low as possible, so most of the quantum signal will be used for generating secure keys.

The polarization of the $\lambda_1$ and $\lambda_2$ light is the same at the input of the fiber channel. However, due to the wavelength dependent birefringence in single-mode fiber, the polarization of the two wavelengths might be different at Charlie. So simply controlling the polarization of the $\lambda_1$ wavelength may result an arbitrary polarization in $\lambda_2$. Here we developed a combined polarization control method to fulfill the requirement of controlling the polarization of both wavelengths.

The polarization control method composes of the following steps. First, the polarization of the $\lambda_1$ is adjusted to an target value, e.g., 100 kHz, are detected at the monitor port. In this procedure, a simple polarization feedback method by attempting each paddle is used. An expected range of the $\lambda_1$ count rates, e.g., between 75 kHz and 300 kHz, is also set. After the polarization reaches the expected value, the second step is started, to minimize the detected count rate of $\lambda_2$ at the monitor port. In this step, the simple polarization feedback method is used with the $\lambda_2$ detected count rate as the input. The polarization drift of $\lambda_1$ is omitted if the $\lambda_1$ count rate falls within the expected range. If the $\lambda_1$ count rate than the upper limit of the expect range, the first step starts again to adjust the $\lambda_1$ count rate to the expected value. By repeating the two steps alternatively, the count rates of both wavelengths will approach the preset range. When the count rate of $\lambda_2$ light reaches the target value, e.g., 100 Hz, the algorithm stops. 
Similarly, an expected range of the $\lambda_2$ count rates, e.g., between 0 Hz and 150 Hz, is set. The algorithm will start again when either the $\lambda_1$ count rate or the the $\lambda_2$ count rate reaches the limit of their expected range. The two-step cycle will stabilize the polarization to within the preset ranges. We note that in the simple polarization feedback method, a smaller voltage step is set when the count rate approaches the target value, for precise adjustion.

\begin{figure}[tbh]
\centering
\resizebox{8.5cm}{!}
{\includegraphics{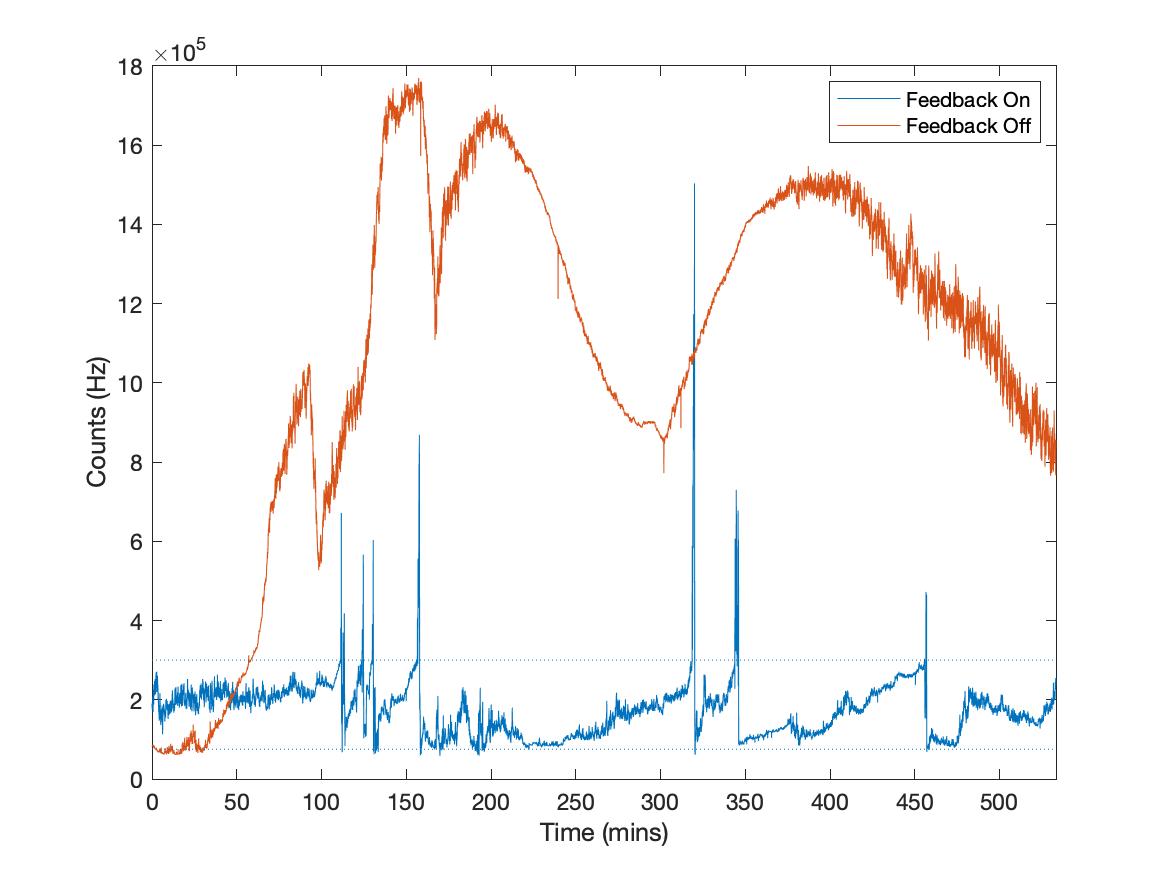}}
\caption{Measured polarization drift of $\lambda_1$ ``strong phase reference'', with and without the polarization feedback. The dashed line indicated the expected range [75 kHz, 300 kHz]. The count rate of $\lambda_1$ falls in the preset range in 98\% of the time, when the feedback is on.}
\label{Fig:SM:PolDriftLambda1}
\end{figure}

\begin{figure}[tbh]
\centering
\resizebox{8.5cm}{!}
{\includegraphics{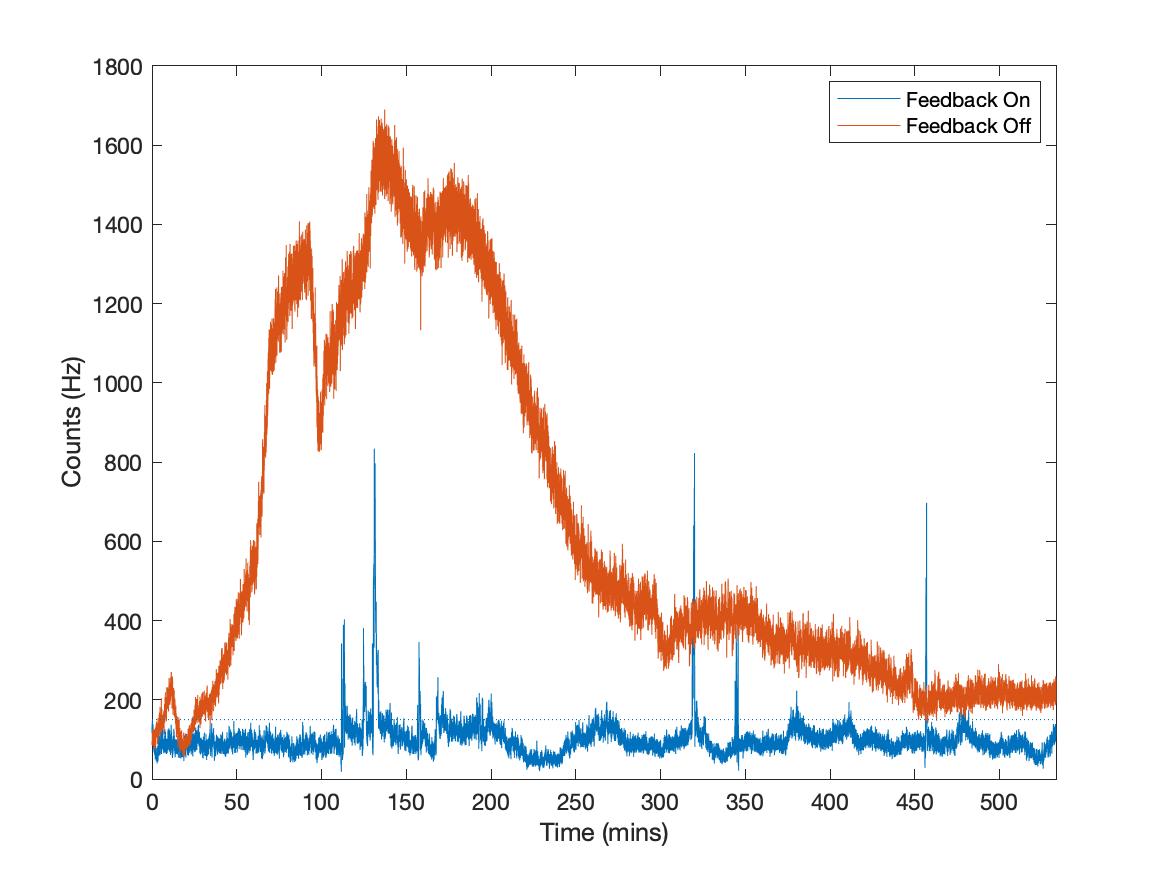}}
\caption{Measured polarization drift of $\lambda_2$ ``dim phase reference'', with and without the polarization feedback. The dashed line indicated the expected range of [0 Hz, 150 Hz]. The count rate of $\lambda_2$ falls in the preset range in 95\% of the time, when the feedback is on.}
\label{Fig:SM:PolDriftLambda2}
\end{figure}

The result of 9 hours polarization control is shown in Fig~\ref{Fig:SM:PolDriftLambda1} and Fig~\ref{Fig:SM:PolDriftLambda2}. A comparison test during the same time span is also performed and shown in the figure. When the polarization control is on, the count rate of the $\lambda_1$ light falls in the range of 75 kHz to 300 kHz in 98\% of the time; the count rate of the $\lambda_2$ light is below 150 Hz in 95\% of the time. As a comparison, when polarization control is off, the count rate of the $\lambda_1$ light drifts to higher than 1 MHz within 2 hours; the polarization drifts to almost the orthogonal state in about 2.5 hours. The count rate of the $\lambda_2$ light drifts out of the preset range in a few minutes; the polarization drifts to almost the orthogonal state in 2.5 hours. The result shows the polarization feedback system works most of the time. We expect to further improve the performance in the future.

\subsection{Local Intensity Feedback}

The bias of Alice's (Bob's) intensity modulators may drift during the experiment. One solution is to monitor the output of the modulators and to adjust the bias based on the measured result. In our experimental setup, a beam splitter is inserted in the signal modulation path for monitoring and feedback. For this monitor, a fraction of 10\% strong light is divided, attenuated, and detected with an SNSPD. The detection result is resorted with the random number used for modulation, to calculate the intensity of the ``dim phase reference'' and the decoy intensities including $\mu_y$, $\mu_x$, and vacuum state. With this monitor, it is convenient to stabilize the intensities by adjusting the bias voltage of the intensity modulators.

Here we use the ratio between the ``dim phase reference'' and the signal states, and that between the $\mu_y$ and $\mu_x$ decoy states, as the input of the feedback system. We adopted a PID algorithm to stabilize the values, by adjusting the bias voltages of the intensity modulators that generating the ``dim phase reference'' pulse and modulating the decoy states. 

\begin{figure}[tbh]
\centering
\resizebox{8.5cm}{!}
{\includegraphics{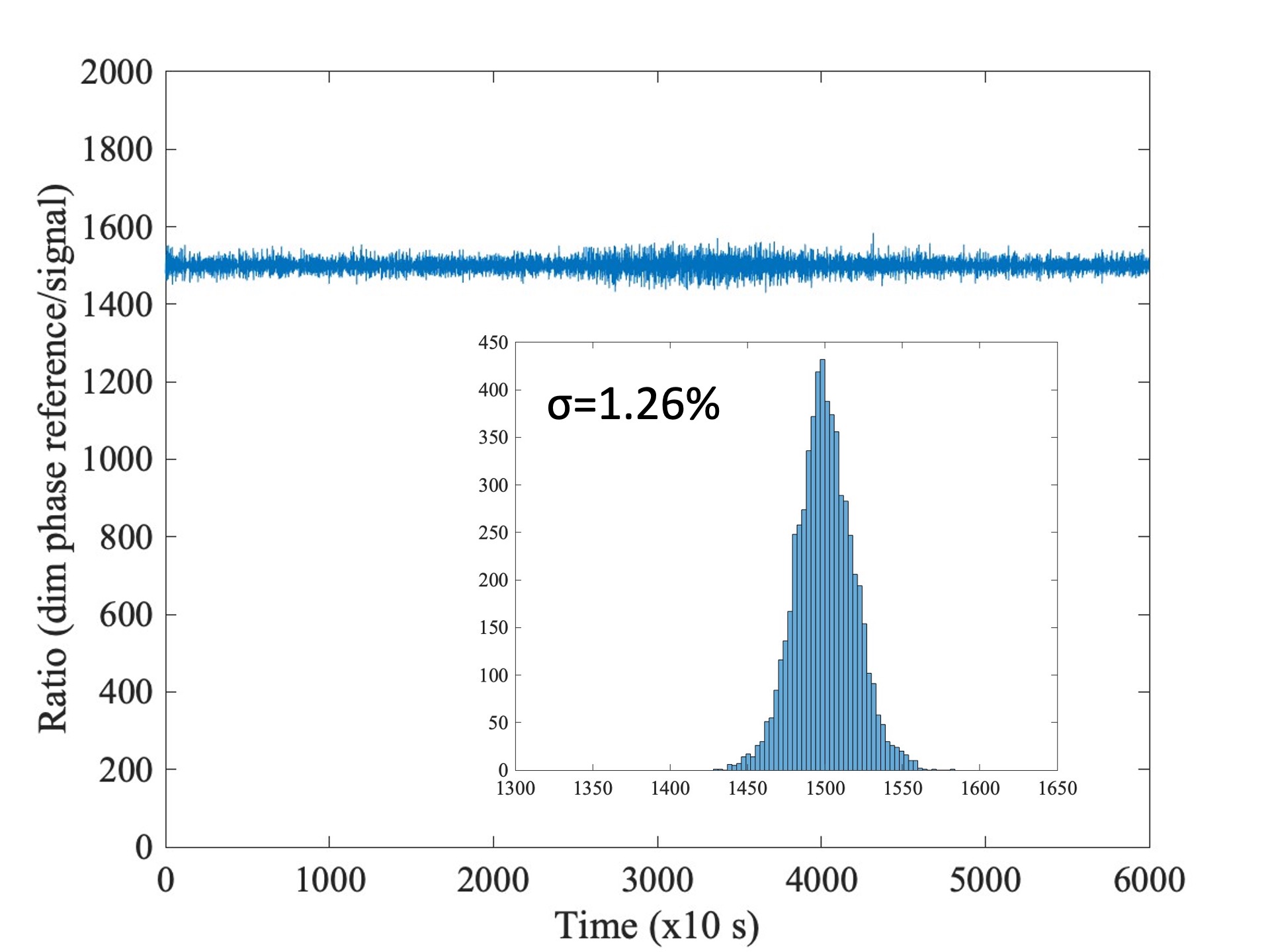}}
\caption{Measured ratio between the ``dim phase reference'' and signal states. The relative error of the ratio is measured to be 1.26\%.}
\label{Fig:SM:RatioSR}
\end{figure}

\begin{figure}[tbh]
\centering
\resizebox{8.5cm}{!}
{\includegraphics{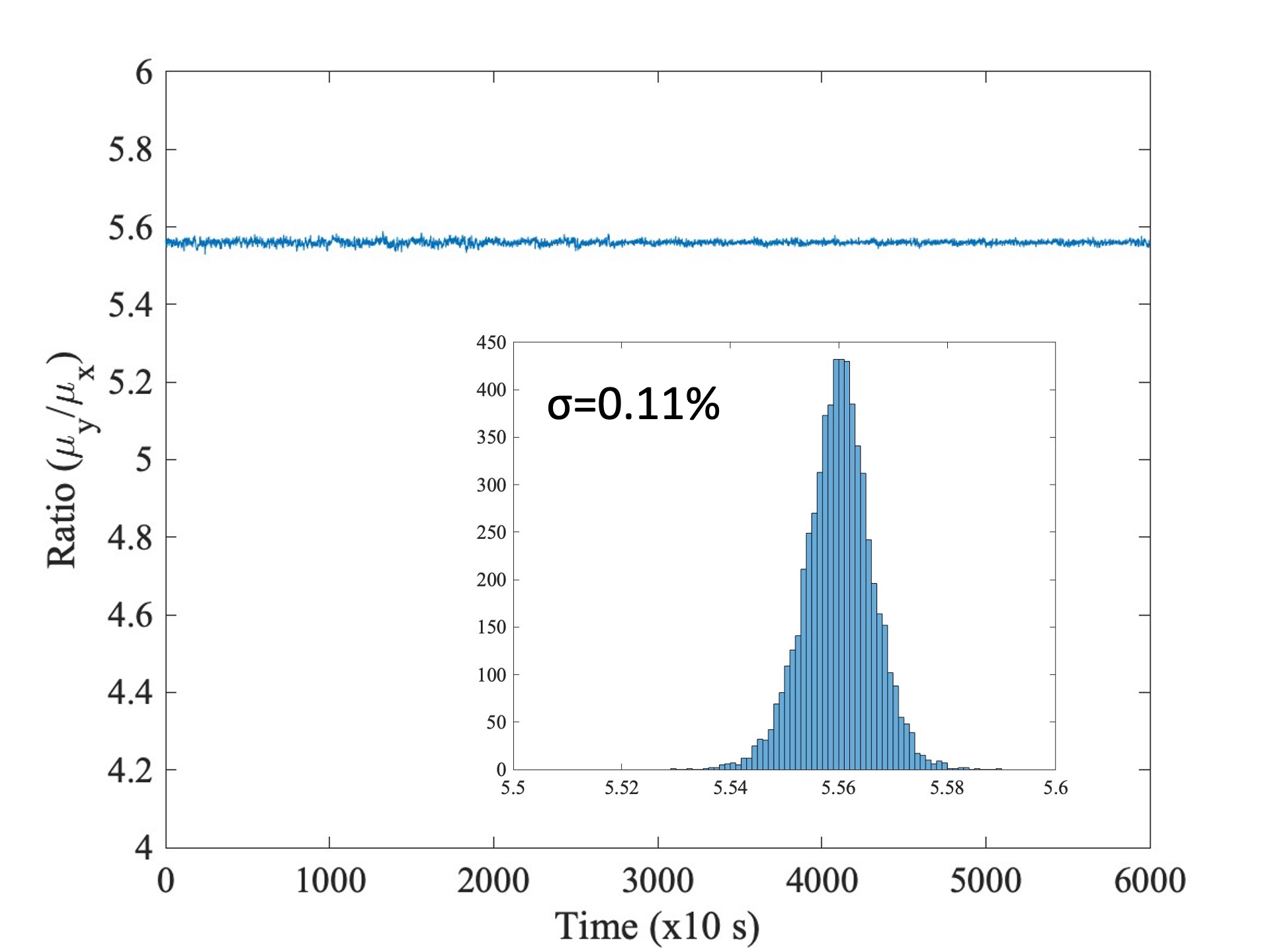}}
\caption{Measured ratio between the $\mu_y$ and $\mu_x$ decoy states. The relative error of the ratio is measured to be 0.11\%.}
\label{Fig:SM:RatioDecoy}
\end{figure}

One test result of around 17 hours is shown in Fig.~\ref{Fig:SM:RatioSR} and Fig.~\ref{Fig:SM:RatioDecoy}. The relative error of the ratio between the ``dim phase reference'' and signal states is measured as 1.26\%; The relative error of the ratio between the $\mu_y$ and $\mu_x$ decoy states is measured as 0.11\%. Thus, the long term stability is guaranteed with the feedback.

\section{Detailed Experimental Results}

\subsection{Detailed Experimental Parameters}

\begin{table*}[htb]
\centering
  \caption{Experimental parameters optimized for different scenarios.}
\begin{tabular}{c|cc|cccc}
\hline
Optimization Condition & $\mu_x$ & $\mu_y$ & $p_{vac}$ & $p_{x}$ & $p_{y}$\\
for Long Distance & 0.08 & 0.445 & 0.52 & 0.28 & 0.20 \\
for Short Distance & 0.08 & 0.445 & 0.68 & 0.04 & 0.28 \\
\hline
\end{tabular}
\label{Tab:parameters}
\end{table*}

In the experiment, we used two sets of parameters. One set is optimized for long distances, the other is for short distances. The parameters are shown in Tab.~\ref{Tab:parameters}, where the ratio of sending the vacuum, $x$ and $y$ states are $p_{vac}$, $p_x$, and $p_y$. The intensity of the $x$ and $y$ states are $\mu_x$ and $\mu_y$ respectively.

\subsection{Characterization of the Channel losses}

\begin{table*}[htb]
\centering
\caption{Experimental characterization of the fibers.}
\begin{tabular}{c|ccccc}
\hline
Total Fiber Length	& 202 km   & 297 km     & 398 km      & 499 km      & 600 km\\
$L_{A-C}$		& 101.0 km   & 148.7 km   & 199.2 km   & 249.7 km   & 300.2 km   \\
$L_{B-C}$		& 101.0 km   & 147.8 km   & 198.3 km   & 248.8 km   & 299.3 km   \\
$\eta_{A-C}$	& 15.8 dB   & 23.3 dB     & 31.2 dB      & 39.1dB       & 47.0 dB     \\
$\eta_{B-C}$	& 15.8 dB   & 22.9 dB     & 30.8 dB       & 38.7dB       & 46.7dB      \\
\hline
Total Fiber Length	    & 701 km      & 800 km     & 901 km      & 952 km      & 1002 km   \\
$L_{A-C}$		  & 350.7 km   & 399.1 km   & 449.6 km   & 474.8 km   & 500.1 km   \\
$L_{B-C}$		 & 348.8 km   & 401.1 km   & 451.6 km    & 476.8 km  & 502.1 km   \\
$\eta_{A-C}$	    & 54.9 dB      & 62.5 dB     & 70.4 dB      & 74.4 dB      & 78.3 dB     \\
$\eta_{B-C}$	    & 54.5 dB      & 62.4 dB     & 70.3 dB       & 74.3 dB     & 78.2 dB     \\
\hline
\end{tabular}
\label{Tab:Characterization}
\end{table*}

The fiber between Alice and Charlie is symmetrical, but the fiber length is not exactly the same. The detailed fiber distances and losses are listed in Tab.~\ref{Tab:Characterization}, where $L_{A-C}$  ($L_{B-C}$) is the fiber distance between Alice (Bob) and Charlie, $\eta_{A-C}$ ($\eta_{B-C}$) is the attenuation between Alice (Bob) and Charlie.

\subsection{Detailed Experimental Results}

The experimental results are summarized in Tab.~\ref{Tab:Result1}, Tab.~\ref{Tab:Result2}, Tab.~\ref{Tab:Result3} and Tab.~\ref{Tab:Result4}.
In the table, we denote $N_{total}$ as the total number of signal pulses, $n_t$(After AOPP) as the remaining pairs after AOPP, $n_1$(Before AOPP) ($n_1$(After AOPP)) as the number of the untagged bits, $e_1^{ph}$(Before AOPP) ($e_1^{ph}$(After AOPP)) as the phase-flip error rate, and QBER E(BeforeAOPP)/E(After AOPP) as the bit-flip error rate before/after the bit error rejection by active odd parity pairing (AOPP). With all the parameters in the table, the final key rate per pulse and in one second is calculated as $R$ (per pulse) and $R$ (bps). We note that the ultra-low QBER E(After AOPP) allows us to use a practical error correction inefficiency $f=1.16$ in calculating the secure key rate.

In calculation, the chosen phase difference is selected as $Ds$ (in degrees). In data processing, we set a detection window to filter noises.  The fraction of the data falling within the detection window is calculated as $r_{window}$. In the cases using long-distance optimized parameters, we set the window width to 200 ps,  $r_{window}\approx$65\% of the detections are kept as valid detection. In cases using short-distance optimized parameters, we set the window width to 400 ps to 500 ps, $r_{window}\approx$90\% of the detections are kept as valid detection.

In the following rows, we list the numbers of pulses Alice and Bob sent in different decoy states, labelled as ``Sent-AB'', where ``A'' (``B'') is ``0'', ``1'', or  ``2'', indicating the intensity Alice (Bob) has chosen within ``vacuum'', ``$\mu_x$'', or ``$\mu_y$''. With the same rule, the numbers of detections are listed as ``Detected-AB''. The total valid detections reported by Charlie is denoted as ``Detected-Valid-ch'', where ``ch'' can be ``Det1'' or ``Det2'' indicating the responsive detector of the recorded counts. The valid events falls in Ds angle range is denoted as ``Detected-11-Ds'', the numbers of correct detections in this Ds range is denoted as ``Correct-11-Ds''.

\linespread{1.2} 

\begin{table*}[htb]
\centering
  \caption{Experimental results for fiber lengths between 499 km and 800 km. All the experimental results shown in this table are tested with the parameters optimized for long distance.}
\begin{tabular}{c|cccc}
\hline
Fiber Length 			& 499 km 		& 600 km 		& 701 km 		& 800 km\\
\hline
$N_{total}$				& $1.264\times 10^{12}$ 	& $1.516\times 10^{13}$       	& $1.516\times 10^{13}$      	& $1.516\times 10^{13}$\\
$R$ (per pulse)   		 		& $1.37\times10^{-7}$        	& $2.28\times10^{-8}$ 	& $2.79\times10^{-9}$         	& $1.46\times10^{-10}$\\
$R$ (bps)			& 47.9 & 7.99 & 0.98 & 0.051 \\
\hline
$n_1$(Before AOPP)		& 2153470                              & 3710700                     	& 570649                      	& 99611 \\
$n_1$(After AOPP)		                & 364683                                & 632284                       	& 96102                        	& 14327 \\
$e_{11}^{ph}$(Before AOPP)                   & 5.01\%                                & 4.31\%                       	& 5.28\%                        	& 7.45\%\\
$e_{11}^{ph}$(After AOPP)                     & 10.56\%                              	& 9.00\%                       	& 12.21\%                      	& 21.44\%\\
E(Before AOPP)			& 27.13\%                               & 28.00\%                     	& 27.97\%                      	& 27.14\%\\
E(After AOPP) 	 		& $1.09\times10^{-3}$	& $4.71\times10^{-4}$   	& $5.20\times10^{-4}$  	& $1.62\times10^{-3}$\\
QBER(X11)	                                & 3.47\%                                 & 3.18\%                         	& 3.40\%                       	& 3.41\%\\
$Ds$		                                &$12^\circ$    	                &$12^\circ$    	        	&$15^\circ$   		&$15^\circ$\\
\hline
Sent-00			& 339697800000		& 4116808800000		& 4116808800000		& 4076373600000\\
Sent-01			& 184906800000		& 2201191200000	                & 2201191200000	                & 2218881600000\\
Sent-10			& 185117400000		& 2198664000000 		& 2198664000000	                & 2221408800000\\
Sent-02			& 132467400000		& 1566864000000 		& 1566864000000 		& 1589608800000\\
Sent-20			& 132256800000		& 1569391200000	                & 1569391200000	                & 1587081600000\\
Sent-12			& 70129800000		& 851666400000	                & 851666400000	    	& 841557600000	\\
Sent-21			& 70340400000		& 849139200000 	                & 849139200000	    	& 844084800000	\\
Sent-11			& 98560800000		& 1195365600000 		& 1195365600000	    	& 1182729600000\\
Sent-22			& 50122800000		& 614109600000	                & 614109600000	    	& 601473600000\\
\hline
Detected-Valid-Det1		& 4423970 		& 7508051		& 1158255		& 222200\\	
Detected-Valid-Det2		& 4050288 		& 6700582		& 1084655		& 200483\\
\hline
Detected-00			& 2706			& 1748			& 308			& 197\\
Detected-01			& 451209	    & 783702 		& 115990			& 22711\\
Detected-10			& 439178 		& 719400 		& 121046			& 21758\\
Detected-02			& 1802616		& 3063579 		& 453736 			& 90541\\
Detected-20			& 1735400		& 2821869		& 476382			& 85859\\
Detected-12			& 1139595		& 1919326		& 289765			& 57271\\
Detected-21			& 1109942		& 1823878 		& 300331			& 54877\\
Detected-11			& 479375	    & 787827 		& 124410			& 23963\\
Detected-22			& 1314237		& 2287304		& 360942			& 65506\\
\hline
Detected-11-Ds	 		& 66781			& 109632			& 21405			& 4075\\
Correct-11-Ds  			& 64466			& 106151			& 20678			& 3936\\
$n_t$(After AOPP)			& 1005228 		& 1615022 		& 265598 		& 50028\\
\hline
\end{tabular}
\label{Tab:Result1}
\end{table*}

\begin{table*}[htb]
\centering
  \caption{Experimental results for fiber lengths between 901 km and 1002 km. All the experimental results shown in this table are tested with the parameters optimized for long distance.}
\begin{tabular}{c|ccc}
\hline
Fiber Length 		& 901 km 		& 952 km			& 1002 km \\
\hline
$N_{total}$			& $1.024\times10^{14}$       	& $5.181\times10^{14}$	& $1.024\times10^{14}$\\
$R$ (per pulse)   	& $2.23\times10^{-11}$        	& $8.75\times10^{-12}$		& $9.53\times10^{-12}$\\

$R$ (bps)			& 0.0078 & 0.0031 & 0.0034 \\
\hline
$n_1$(Before AOPP)		& 114722		& 249989		& 21725 \\
$n_1$(After AOPP)		& 17162         & 39968			& 3853  \\
$e_{11}^{ph}$(Before AOPP)	& 8.01\% 	& 6.32\%		& 2.84\%\\
$e_{11}^{ph}$(After AOPP)   & 21.88\%   & 15.76\%		& 5.51\%\\
E(Before AOPP)			& 27.18\%       & 28.52\%		& 28.45\%  \\
E(After AOPP) 	 		& $2.15\times10^{-3}$     	& $8.80\times10^{-3}$		& $1.99\times10^{-2}$ \\
QBER(X11)	            & 3.80\%        & 3.58\%		& 3.30\%\\
$Ds$		            &$15^\circ$   	&$12^\circ$		&$10^\circ$\\
\hline
Sent-00				& 27545848200000	   	& 140571288000000	& 27668206800000\\
Sent-01			& 14964183000000    	& 75121020000000	& 14908163400000\\
Sent-10			& 14977450800000	   	& 75293712000000	& 14931329400000\\
Sent-02			& 10712800800000    	& 53707212000000	& 10646461800000\\
Sent-20			& 10699533000000    	& 53534520000000	& 10623295800000\\
Sent-12			& 5688095400000	    	& 28925910000000	& 5733585000000\\
Sent-21			& 5701363200000     	& 29098602000000	& 5756751000000\\
Sent-11			& 7992901800000	    	& 40841658000000	& 7993533600000\\
Sent-22			& 4069423800000	    	& 20982078000000	& 4090273200000\\
\hline
Detected-Valid-Det1		& 247455	 		& 557129	& 44242\\	
Detected-Valid-Det2		& 235749    		& 462342	& 41536\\
\hline
Detected-00			& 335	   		& 2365		& 485	\\
Detected-01			& 24636    		& 53199		& 4430	\\
Detected-10			& 26341	   		& 55148		& 4804	\\
Detected-02			& 99046    		& 204846		& 17053	\\
Detected-20			& 102681   		& 210005		& 18136	\\
Detected-12			& 63022	   		& 132731		& 10782	\\
Detected-21			& 64631	   		& 139240		& 11611	\\
Detected-11			& 27568	   		& 58768			& 4967	\\
Detected-22			& 74944	   		& 163169		& 13510	\\
\hline
Detected-11-Ds	 		& 4840      		& 8162	& 576\\
Correct-11-Ds  			& 4656  			& 7870	& 557\\          
$n_t$(After AOPP)		& 58629 			& 117913	& 10343	\\
\hline
\end{tabular}
\label{Tab:Result2}
\end{table*}

\begin{table*}[htb]
\centering
  \caption{Experimental results for fiber lengths between 202 km and 398 km. Note the experimental results are tested with the parameters optimized for long distance.}
\begin{tabular}{c|ccc}
\hline
Fiber Length 			& 202 km 		& 297 km 		& 398 km \\
\hline
$N_{total}$	 		& $1.264\times 10^{12}$ 	& $1.264\times 10^{12}$ 	& $1.264\times 10^{12}$ \\
$R$ (per pulse)   				& $3.01\times 10^{-5}$ 	& $5.60\times 10^{-6}$ 	& $9.77\times 10^{-7}$ \\
$R$ (bps) 		& 10561.3 & 1966.1 & 343.0 \\
\hline
$n_1$(Before AOPP)			&463719000   		& 82246800  		& 13567700 \\
$n_1$(After AOPP)			&82795500   		& 14316500  		& 2337260 \\
$e_{11}^{ph}$(Before AOPP)  		& 4.30\%  		& 4.10\% 			& 4.39\% \\
$e_{11}^{ph}$(After AOPP)   		& 8.28\%  		& 8.02\% 			& 8.77\% \\
E(Before AOPP)			& 28.45\%  		& 27.14\% 		& 27.09\% \\
E(After AOPP) 	 		& $5.04\times 10^{-3}$ 	& $3.57\times 10^{-3}$ 	 & $1.26\times 10^{-3}$ \\
QBER(X11)	 			& 3.59\%  		&  3.33\%		 &  3.34\% \\
$Ds$		 			& $10^\circ$ 		& $10^\circ$ 	 	& $12^\circ$ \\
\hline
Sent-00			& 342856800000 		& 339697800000  		& 339697800000 \\
Sent-01			& 183222000000		& 184906800000  		& 184906800000 \\
Sent-10			& 183643200000		& 185117400000  		& 185117400000 \\
Sent-02			& 130993200000		& 132467400000  		& 132467400000 \\
Sent-20			& 130572000000		& 132256800000  		& 132256800000 \\
Sent-12			& 70551000000 		& 70129800000   		& 70129800000 \\
Sent-21			& 70972200000  		& 70340400000   		& 70340400000 \\
Sent-11			& 99613800000 		& 98560800000   		& 98560800000 \\
Sent-22			& 51175800000 		& 50122800000   		& 50122800000 \\
\hline
Detected-Valid-Det1		& 985627697  		& 166652028  		& 26951345 \\
Detected-Valid-Det2		& 784707910  		& 150869935  		& 25540660 \\
\hline
Detected-00			&2311526   		& 318619			& 18551 \\
Detected-01			&94429685   		& 17353062  		& 2869579 \\
Detected-10			&92544595   		& 16252690  		& 2665399 \\
Detected-02			&364017816   		& 68601849  		& 11380162 \\
Detected-20			&356937380   		& 63800927  		& 10531056 \\
Detected-12			&237189175   		& 43122756  		& 7146935  \\
Detected-21			&236946702   		& 41010260  		& 6778067 \\
Detected-11			&101640635   		& 18052870  		& 2978564 \\
Detected-22			&284318093   		& 49008930  		& 8123692 \\
\hline
Detected-11-Ds 			&11851523   		& 2107814  		& 414620  \\
Correct-11-Ds  			&11425566   		& 2037601  		& 400786  \\
$n_t$(After AOPP)		& 201229804   		& 37134329  	& 6153487 \\
\hline
\end{tabular}
\label{Tab:Result3}
\end{table*}

\begin{table*}[htb]
\centering
  \caption{Experimental results for fiber lengths between 202 km and 398 km. Note that the experimental results with ``*'' in this table are test with the parameters optimized for short distance.}
\begin{tabular}{c|cccccc}
\hline
Fiber Length 			& 202 km* 		& 297 km* 		& 398 km* \\
\hline
$N_{total}$	 		& $3.240\times 10^{12}$ 	& $3.240\times 10^{12}$  	& $3.240\times 10^{12}$\\
$R$ (per pulse)   	& $5.23\times 10^{-5} $  	& $1.26\times 10^{-5} $  	& $2.38\times 10^{-6} $\\
$R$ (bps) 		& 47061.8 & 11327.8 & 2140.7\\
\hline
$n_1$(Before AOPP)		& 2934440000  		& 584003000  		& 93786900  \\
$n_1$(After AOPP)		& 511247000   		& 102295000  		& 16114500  \\
$e_{11}^{ph}$(Before AOPP)  		& 5.38 \%  			& 6.75\%   			& 5.46\%\\
$e_{11}^{ph}$(After AOPP)   		& 10.22\%  			& 12.69\%   			& 10.47\%\\
E(Before AOPP)			& 29.28\%  		& 29.18\%   		& 29.18\%\\
E(After AOPP) 	 		& $7.56\times 10^{-3} $  	& $1.69\times 10^{-3} $  	& $1.15\times 10^{-3} $\\
QBER(X11)	 			&  4.21\%  		& 4.83\%   		& 3.49\%\\
$Ds$		 	& $10^\circ$  		& $10^\circ$  		& $12^\circ$\\
\hline
Sent-00			& 1497960000000   	& 1497960000000		& 1497960000000  \\
Sent-01			& 87480000000   		& 87480000000  		& 87480000000  \\
Sent-10			& 88560000000   		& 88560000000 		& 88560000000  \\
Sent-02			& 617760000000   		& 617760000000  		& 617760000000 \\
Sent-20			& 616680000000   		& 616680000000  		& 616680000000 \\
Sent-12			& 35640000000  		& 35640000000		& 35640000000  \\
Sent-21			& 36720000000 		& 36720000000		& 36720000000  \\
Sent-11			& 5400000000  		& 5400000000 		& 5400000000   \\
Sent-22			& 253800000000  		& 253800000000  		& 253800000000 \\
\hline
Detected-Valid-Det1		& 3981108468  		& 735082715  		& 115177244  \\
Detected-Valid-Det2		& 3133795324  		& 675778879  		& 110657343  \\
\hline
Detected-00			& 21840742  		& 957814  		& 103896  \\
Detected-01			& 58994154 		& 11431115  		& 1923344  \\
Detected-10			& 63653547  		& 12639747  	& 1945655  \\
Detected-02			& 2279905052  		& 447932487  		& 74864277  \\
Detected-20			& 2429904373  		& 487302891  		& 74822343  \\
Detected-12			& 156288238  		& 31053547  		& 5123011  \\
Detected-21		& 168270293  		& 33710292  		& 5253612  \\
Detected-11			& 7647482  		& 1477134  		& 237562  \\
Detected-22			& 1928399911  		& 384356567  		& 61560887  \\
\hline
Detected-11-Ds 			& 891843  		& 172509  		& 32943 \\
Correct-11-Ds  			& 854313  		& 164174  		& 31794 \\
$n_t$(After AOPP)		& 1323118973  		& 257696601  		& 41072397  \\
\hline
\end{tabular}
\label{Tab:Result4}
\end{table*}

\bibliography{BibSNSTFQKD}

\end{document}